\documentclass[conference]{IEEEtran}

\usepackage{lineno}
\modulolinenumbers[5]

\usepackage{graphicx}
\usepackage[cmex10]{amsmath}
\usepackage{array}
\usepackage{mdwmath}
\usepackage{mdwtab}
\usepackage{eqparbox}
\usepackage{subfig}
\usepackage{caption}
\usepackage{stfloats}
\usepackage{url}
\usepackage{amsfonts,amssymb}
\usepackage{pifont}
\usepackage{ctable}
\usepackage{multirow}
\usepackage{diagbox}

\usepackage[ruled,vlined]{algorithm2e}

\SetKwComment{Comment}{/* }{ */}
\SetKwInput{KwIn}{In}
\SetKwInput{KwOut}{Out}

\newcommand{\eat}[1]

\newcommand{\circnum}[1]{\ding{\numexpr171+#1\relax}} 

\graphicspath{{eps/}}

\IEEEoverridecommandlockouts

\begin{document}
	
\title{\LARGE RFID-based Real-Time Geriatric Gait Speed Monitoring System: Design, Implementation and Clinical Evaluation}

	\author{\IEEEauthorblockN{Natong Lin\IEEEauthorrefmark{1},
            Jiachen Wang\IEEEauthorrefmark{1},
            Lisa C. Barry\IEEEauthorrefmark{2},
			Song Han\IEEEauthorrefmark{1}
		}
		\IEEEauthorblockA{\IEEEauthorrefmark{1}School of Computing, University of Connecticut E-mails:$\{firstname.lastname\}@uconn.edu$}
        \IEEEauthorblockA{\IEEEauthorrefmark{2}School of Medicine, University of Connecticut E-mails:$libarry@uchc.edu$}

	}

\maketitle

\newcommand{\han}[1] {{\bf \color{red}{(SH: #1)}}}
\newcommand{\lin}[1] {{\bf \color{blue}{#1}}}

\begin{abstract}
Gait speed is a widely used indicator of functional health and mobility decline, yet in clinical practice it is commonly measured manually using a stopwatch, which limits scalability and measurement frequency. Privacy-preserving and maintenance-free sensing approaches can enable more routine and less burdensome assessments in real-world care settings.  This paper presents the design, implementation, and real-world deployment of a fully passive, battery-free gait-speed monitoring system based on ultra-high-frequency (UHF) RFID. Compared with camera- and wearable-based approaches, the proposed system preserves patient privacy by avoiding video capture and biometric data, while eliminating battery maintenance. The system employs a dual-antenna configuration and an edge-based peak-detection algorithm to estimate gait speed in real time from received signal strength indicator (RSSI) streams. By leveraging antenna-beam symmetry and asymmetric signal processing, the method improves robustness to noise, plateau regions, and multiple local maxima.  We evaluate the system during routine outpatient care across three clinical sites using 966 trials, achieving an 87.7\% measurement success rate. Compared with concurrent stopwatch timing, the system attains a mean absolute error of 0.064~$m/s$, demonstrating reliable operation with accuracy suitable for clinical gait-speed assessment.
\end{abstract}

\section{Introduction}
Gait speed is a simple yet powerful indicator of functional health and mobility decline. In older adults, gait speed is strongly associated with survival, functional status, and self-rated health, and has been shown to provide reliable and valid assessment across different walking distances and pacing conditions~\cite{middleton2015walking,studenski2011gait}. Despite its clinical value, gait speed is often measured manually using stopwatch timing, which requires staff attention and typically occurs only during scheduled visits. This limits measurement frequency and may fail to capture typical performance, as patients often walk differently when formally observed than during routine clinical activities.

To reduce manual burden and enable more scalable monitoring, many automated gait measurement systems have been proposed~\cite{mejiacruz2021walking}. Camera-based systems can estimate gait parameters with high accuracy, but they raise privacy concerns and typically require carefully controlled placement~\cite{springer2016validity,mazurek2024validation}. Wearable-based approaches reduce environmental dependencies but impose burdens related to battery charging, device management, and user compliance~\cite{muro2014gait}. Other modalities, such as radar, ultra-wideband (UWB), Bluetooth, and WiFi sensing, have also been explored; however, they often require specialized hardware or site-specific calibration---factors that can limit deployment in routine care settings~\cite{gurbuz2024overview,sansano2022continuous,wu2020gaitway}.

Passive ultra-high-frequency (UHF) RFID provides an attractive alternative that aligns well with clinical deployment constraints. Passive RFID tags are battery-free, low-cost, and already common in healthcare environments for identification and tracking, while preserving privacy by avoiding video capture or biometric recording. Our team previously demonstrated the clinical feasibility of passive RFID for gait speed measurement, showing high agreement with manual stopwatch timing and strong acceptance from both patients and providers~\cite{barry2018design}. More recently, we showed that RFID-based gait speed assessment can be integrated into Medicare Annual Wellness Visit workflows across multiple primary care practices with high implementation fidelity following workflow and system refinements~\cite{barry2025taking}. However, neither study addressed the underlying system design or signal-processing methodology. The earlier work validated clinical acceptability using manual timing as a reference, while the latter focused on workflow adoption rather than the technical architecture enabling real-time, automated measurement. This paper fills that technical gap by presenting a complete system architecture for passive RFID-based gait speed monitoring, including hardware configuration, real-time signal processing, and multi-site clinical deployment.

Realizing this system for real-time clinical use presents several technical challenges across the signal, system, and deployment levels.
\textit{At the signal level}, RSSI measurements from passive tags are inherently noisy and often exhibit plateau regions near peak signal strength as tags pass close to antennas. In addition, RFID readers stream tag reads continuously without signaling when walking events begin or end, and patients may enter, pause, or restart walking attempts unpredictably. These characteristics make simple threshold-based triggering and offline peak selection unreliable for real-time operation.
\textit{At the system level}, the architecture must process high-rate streaming data with low latency and deliver immediate feedback to clinical staff through a responsive interface.
\textit{At the deployment level}, hardware selection and physical installation must accommodate diverse hallway configurations, varying patient heights, and existing clinical infrastructure across sites, using components that are cost-effective and maintainable without specialized technical support.

We address these challenges through three integrated design decisions:
(i) a dual-antenna, edge-based peak-detection algorithm that exploits antenna beam symmetry to detect consistent edge events rather than ambiguous plateau midpoints, resolving multiple-peak and unknown-termination challenges in streaming RSSI data;
(ii) a Go-based backend that processes tag reads online and pushes results immediately to a browser-based clinical interface via WebSocket, eliminating manual timing and enabling real-time feedback; and
(iii) a hardware configuration built entirely from commercial off-the-shelf RFID components with configurable antenna spacing to support installation across diverse clinical sites.

We validate the system through deployment across three outpatient clinical sites, collecting 966 gait-speed trials during routine care. Across all sites, 87.7\% of trials (847 of 966) produce valid measurements. In a separate comparison against concurrent stopwatch timing, the system achieves a mean absolute error of 0.064~$m/s$ confirming accuracy sufficient for clinical gait speed assessment. All data collection was conducted under institutional review board approval (UConn Health IRB 25-1890-2).
 
The remainder of this paper is organized as follows. Section~\ref{sec:relatedWork} reviews related work on automated gait measurement systems. Section~\ref{sec:sysDesign} describes the proposed RFID-based system, including the hardware configuration and software architecture. Section~\ref{sec:algorithm} presents the real-time, dual-antenna peak-detection algorithm. Section~\ref{sec:deployment} reports results from multi-site clinical deployment and evaluation. Section~\ref{sec:conclusion} concludes the paper and points out the future directions. 

\section{Related Work} \label{sec:relatedWork}
Despite the established clinical value of gait speed as an indicator of functional health, mobility decline, and mortality risk in older adults~\cite{middleton2015walking,studenski2011gait}, accurately measuring gait in real-world settings presents significant technical challenges. Numerous gait measurement technologies have been developed~\cite{mejiacruz2021walking}, employing diverse sensing modalities and signal processing techniques to capture gait parameters. These technologies span a broad spectrum, ranging from contact-based force sensors and wearable devices to non-contact approaches based on computer vision, radio-frequency sensing, and passive infrared detection.  

This section examines the technical characteristics of existing gait measurement approaches, including their measurement accuracy and their capabilities for real-time, automated processing to support integration into clinical workflows. 

Pressure-sensitive walkway systems are among the most widely deployed automated gait measurement technologies in clinical and research settings and have been shown to provide highly accurate and reliable spatiotemporal gait measurements. The GAITRite system employs a pressure-sensitive sensor array to capture spatial and temporal gait parameters with high precision~\cite{mcdonough2001validity} and has been validated in large-scale clinical studies~\cite{briggs2020relationship}. Protokinetics ZenoMat walkways are also widely used in research and are commonly treated as a standard reference in gait analysis studies~\cite{boettcher2020dual,lynall2017reliability,pham2024effects}. Similarly, pressure-sensitive mats such as the Tekscan HR Mat have been used in combination with wireless timing gates to measure plantar pressure distribution at different gait speeds~\cite{kirmizi2020effects}.  While pressure-sensitive walkways provide highly accurate gait measurements, their high cost and reliance on dedicated floor-mounted hardware limit deployment flexibility and scalability, particularly in primary care settings.

To improve portability and reduce infrastructure requirements, many gait assessment approaches rely on wearable sensors that can be easily deployed across diverse environments without dedicated installation~\cite{muro2014gait}. The most common wearable sensors for gait speed measurement are inertial measurement units (IMUs), which combine accelerometers and gyroscopes to capture body motion and can accurately estimate walking speed~\cite{kitagawa2016estimation,washabaugh2017validity,zhou2018hemodialysis,keppler2019validity}. Personalized wrist-mounted IMU models have demonstrated accuracy comparable to global navigation satellite systems (GNSS) (e.g., 0.05~$m/s$ RMSE) while enabling continuous indoor and outdoor monitoring~\cite{soltani2019real}. Insole-based IMU systems leverage foot-mounted sensors to capture both inertial measurements and plantar pressure distribution~\cite{farid2021feetme}. While wearable IMUs demonstrate clinical-grade accuracy, their deployment requires device provisioning, patient training, and ongoing maintenance (e.g., charging and updates), which can limit scalability in primary care and routine outpatient settings.

To avoid body-worn sensors, vision-based approaches using depth cameras such as Microsoft Kinect~\cite{springer2016validity} capture spatiotemporal gait parameters via skeleton-based tracking. Multi-camera Kinect systems validated against VICON motion capture demonstrate excellent agreement for walking speed, stride length, and cadence, with intraclass correlation coefficients exceeding 0.88~\cite{geerse2015kinematic}. LiDAR-based systems using laser ranging show strong agreement with stopwatch measurements for gait speed across 4-m walking tests~\cite{pua2025gait}. Recent portable camera systems similarly report strong correlations ($r > 0.9$) with pressure-sensitive walkways for gait velocity~\cite{mazurek2024validation}. While vision-based systems enable accurate automated measurement without patient-worn devices, they require unobstructed line of sight and controlled lighting, which constrains integration into routine clinical workflows.

In contrast to the above approaches, radio-frequency (RF) sensing has also been used to estimate the speed of moving objects. WiFi channel state information (CSI) provides rich signal features from which gait velocity can be inferred, including Doppler frequency shifts~\cite{li2024wifi,li2017indotrack,qian2017widar,qian2018widar2}, subcarrier phase variations~\cite{li2022wivelo}, and statistical CSI properties~\cite{wu2020gaitway}. Beyond WiFi, Bluetooth Low Energy (BLE) has emerged as an alternative RF sensing modality, offering advantages in power efficiency and integration with personal wearable devices~\cite{sansano2022continuous}. Millimeter-wave radar systems have also been explored for gait analysis, leveraging micro-Doppler signatures to extract biomechanical parameters~\cite{gurbuz2024overview,wang2024extraction}. While radar-based approaches can operate through walls and under non-line-of-sight conditions, they typically require specialized hardware, which limits deployment in routine clinical settings.

While the RF sensing approaches described above demonstrate technical feasibility for contactless gait measurement, their clinical deployment is often hindered by practical challenges, including interference from existing medical equipment in clinical environments. Radio-frequency identification (RFID) offers an alternative RF-based sensing modality that alleviates some of these deployment limitations. RFID operates via passive backscatter, in which battery-free tags costing only cents reflect modulated signals from a reader, enabling sensing without per-patient power management. Prior work has exploited Doppler shifts in backscattered RFID signals to estimate walking speed, achieving approximately 4\% error using a geometric model relating dual-antenna Doppler frequency to tag velocity~\cite{huang2022rfid}. RFID readers have also been deployed at stairwell landings to track indoor physical activity, demonstrating strong correlation ($r = 0.78$–$0.94$) with accelerometer-derived energy expenditure when participants wore passive tags at multiple body locations~\cite{gay2021novel}.

\begin{figure*}[t]
	\centering
	\includegraphics[width = 0.9 \linewidth]{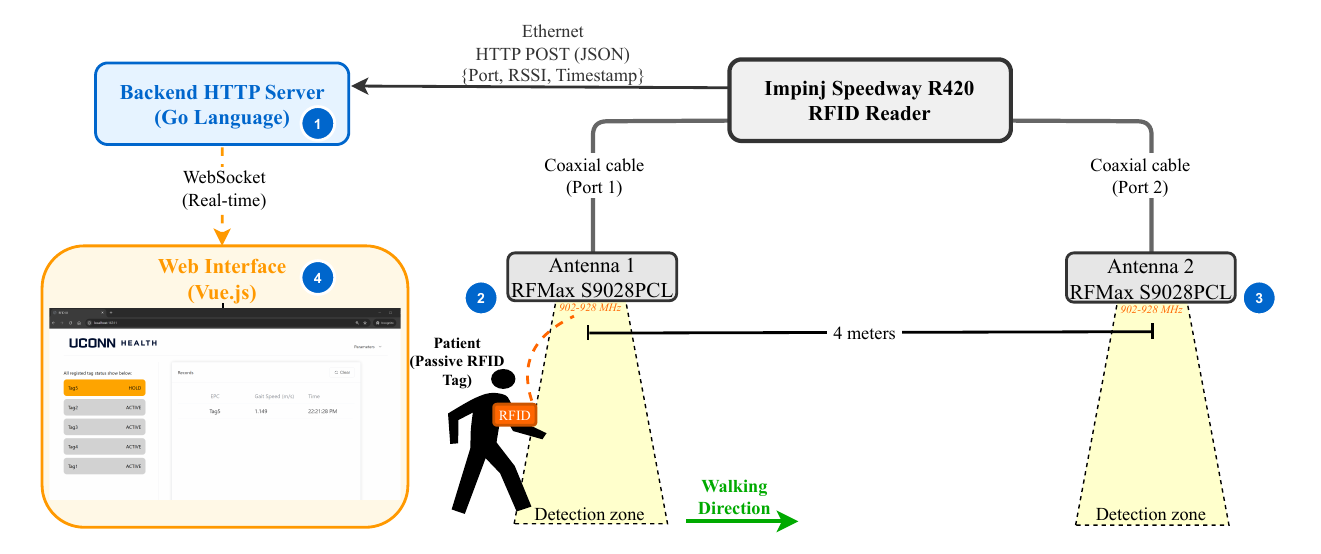}
	\caption{\small High-level system architecture for RFID-based gait-speed monitoring: \circnum{1} the backend server is started and connected to the RFID reader and web UI; \circnum{2} a patient wearing a passive tag walks past the first antenna; \circnum{3} the patient continues past the second antenna while RSSI streams are processed online to estimate gait speed; and \circnum{4} the computed gait speed is reported immediately in the web UI.}
	\label{fig:architecture}
	\vspace{-1em}
\end{figure*}

Beyond feasibility, RFID offers several structural advantages for clinical deployment. Unlike tag-free RF, radar, or vision-based approaches, RFID explicitly associates measurements with individual patients via unique tag identifiers, eliminating multi-target tracking and re-identification challenges in shared clinical spaces. Prior work in a clinical setting similar to ours used dual UHF RFID antennas to estimate average walking speed based on RSSI peak detection and cross-correlation processing, achieving less than 1\% error under controlled laboratory conditions~\cite{mo2018study}.  However, the proposed multi-stage signal processing pipeline requires complete signal availability before processing and does not address real-time streaming constraints where signal termination is unknown, concurrent multi-patient tracking, or integration into routine clinical workflows across multiple sites.

In this work, we address these operational challenges with a dual-antenna, edge-based peak detection algorithm designed for real-time streaming RSSI data, combined with a concurrent processing architecture that supports simultaneous multi-patient tracking. We validate the system through deployment across three clinical sites with 966 trials collected during routine care, demonstrating that RFID-based gait speed monitoring achieves clinical-grade accuracy while integrating seamlessly into existing workflows.

\section{System Design} \label{sec:sysDesign}

\subsection{System Overview}
The proposed RFID-based gait speed measurement system is designed for routine clinical use and employs a dual-antenna configuration to detect patient movement from received signal strength indicator (RSSI) patterns. The system comprises two RFID antennas mounted along one side of a walking path, a commercial RFID reader, a backend processing server, and a web-based user interface. 

An end-to-end system architecture of the proposed gait-speed monitoring system is shown in Fig.~\ref{fig:architecture}. Two spatially separated UHF RFID antennas are connected to a commercial RFID reader, which continuously scans passive tags worn by patients as they walk through the measurement zone. The reader streams timestamped tag-read events, including antenna port identifiers and RSSI values, to a backend processing server over HTTP. The backend server performs online signal processing and gait-speed computation using a concurrent, tag-centric architecture, enabling simultaneous tracking of multiple patients without blocking. Once a gait-speed measurement is completed, the result is immediately pushed to a browser-based user interface via WebSocket, allowing clinical staff to receive feedback with minimal latency. This design avoids offline processing, eliminates manual intervention, and supports seamless integration into routine clinical workflows.

The gait-speed computation exploits the characteristic behavior of RSSI signals as RFID tags move through space. As a patient wearing a passive RFID tag on an armband walks past each antenna, RSSI values increase during approach, reach a peak near the point of closest proximity, and then decrease as the patient moves away. The dual-antenna peak-detection algorithm identifies these peak events to estimate entry and exit timestamps, enabling gait-speed calculation from the known antenna separation and the measured time interval. Unlike simple threshold-based triggering, which is sensitive to noise and transient fluctuations, the proposed moving-window peak-detection approach provides robust measurements across patients with different walking speeds and under varying environmental conditions.

Overall, the system design emphasizes clinical practicality through simple operation (armbands rather than specialized wearables), real-time automated processing (eliminating manual timing), easy installation (configuration-based deployment), and cost-effectiveness using commercial off-the-shelf (COTS) RFID components. The following subsections detail the hardware configuration (Section III.\ref{subsubsec:hardware}) and software architecture (Section III.\ref{subsubsec:software}).

\subsection{Hardware Components}\label{subsubsec:hardware}
The proposed gait-speed measurement system is built entirely from commercial RFID components to support reliable, low-cost clinical deployment. The hardware setup includes two UHF RFID antennas, a commercial RFID reader, passive RFID tags worn by patients, and a PC that runs the backend processing server and hosts the user interface. 

\vspace{0.05in}
\noindent \textbf{RFID Reader.}
The system employs an Impinj Speedway Revolution R420, a 4-port UHF RFID reader designed for fixed infrastructure deployments. The reader operates in the 902–928~MHz frequency band (FCC region) with configurable RF output power. Under optimal conditions, it achieves aggregate read rates of up to 1{,}100 tags per second across all ports and provides microsecond-precision timestamps for each tag-detection event. The reader connects to the backend server via Gigabit Ethernet, enabling straightforward network integration, and is powered through an AC–DC power supply.

The reader is configured using Speedway Connect software with settings optimized for gait measurement in a clinical environment. Reader modes define the over-the-air communication between the reader and tags, trading off read rate against robustness to RF interference through different modulation and encoding schemes~\cite{impinj_reader_modes_2022}. We use Reader Mode~4, which employs higher-order Miller encoding to improve robustness in environments with potential RF interference from other wireless systems common in clinical settings. Although this mode sacrifices peak read rate, the tradeoff is appropriate for gait measurement, where reliable detection of individual patients is prioritized over high-throughput inventory scanning.

The reader operates with Session~1 inventory configuration and Dual-Target search mode. Session~1 is well suited for tracking moving objects~\cite{epc_standard}, as it allows the reader to detect each tag multiple times as patients traverse the measurement zone. As patients walk along the 4~m hallway segment at typical gait speeds (0.6–1.5~$m/s$), this configuration yields approximately 50–150 RSSI samples per antenna, enabling robust peak detection despite noise and multipath interference common in clinical environments. The microsecond-precision timestamps (reported via HTTP POST in JSON format) enable precise gait-speed computation.

\vspace{0.05in}
\noindent \textbf{Antenna.}
Two RFMax S9028PCL antennas (8.5~dBic gain, left-hand circular polarization, 70$^\circ$ beamwidth~\cite{vulcan_s9028pc_antenna}) provide coverage at the entry and exit points of the measurement zone. Circular polarization ensures consistent tag detection regardless of armband orientation. The antennas operate across the 902–928~MHz UHF RFID band and are mounted vertically at a height of 1.2~m, positioned perpendicular to the walking path to maximize RSSI sensitivity. The antenna spacing is set to 4.0~m, providing sufficient distance for clinically meaningful gait assessment; this distance is configurable to accommodate different clinical layouts.

\vspace{0.05in}
\noindent \textbf{RFID Tags.}
Patients wear passive UHF RFID tags compliant with the Electronic Product Code (EPC) Generation~2 (ISO~18000-6C) protocol~\cite{epc_standard}. Passive tags require no battery; instead, they harvest energy from the reader’s RF field to power their circuitry and backscatter responses. Each tag contains a 96-bit EPC memory storing a unique identifier. This passive design eliminates battery maintenance and supports effectively indefinite tag lifetime at minimal cost.

Tags are held in adjustable polyvinyl chloride (PVC) armbands secured with elastic straps, allowing placement on the patient’s upper arm. Upper-arm placement positions the tag near chest height during walking, providing consistent line-of-sight to both antennas across varying patient heights and reducing RSSI fluctuations due to body shadowing effects.

\begin{figure}[t]
	\centering
	\includegraphics[width=\linewidth]{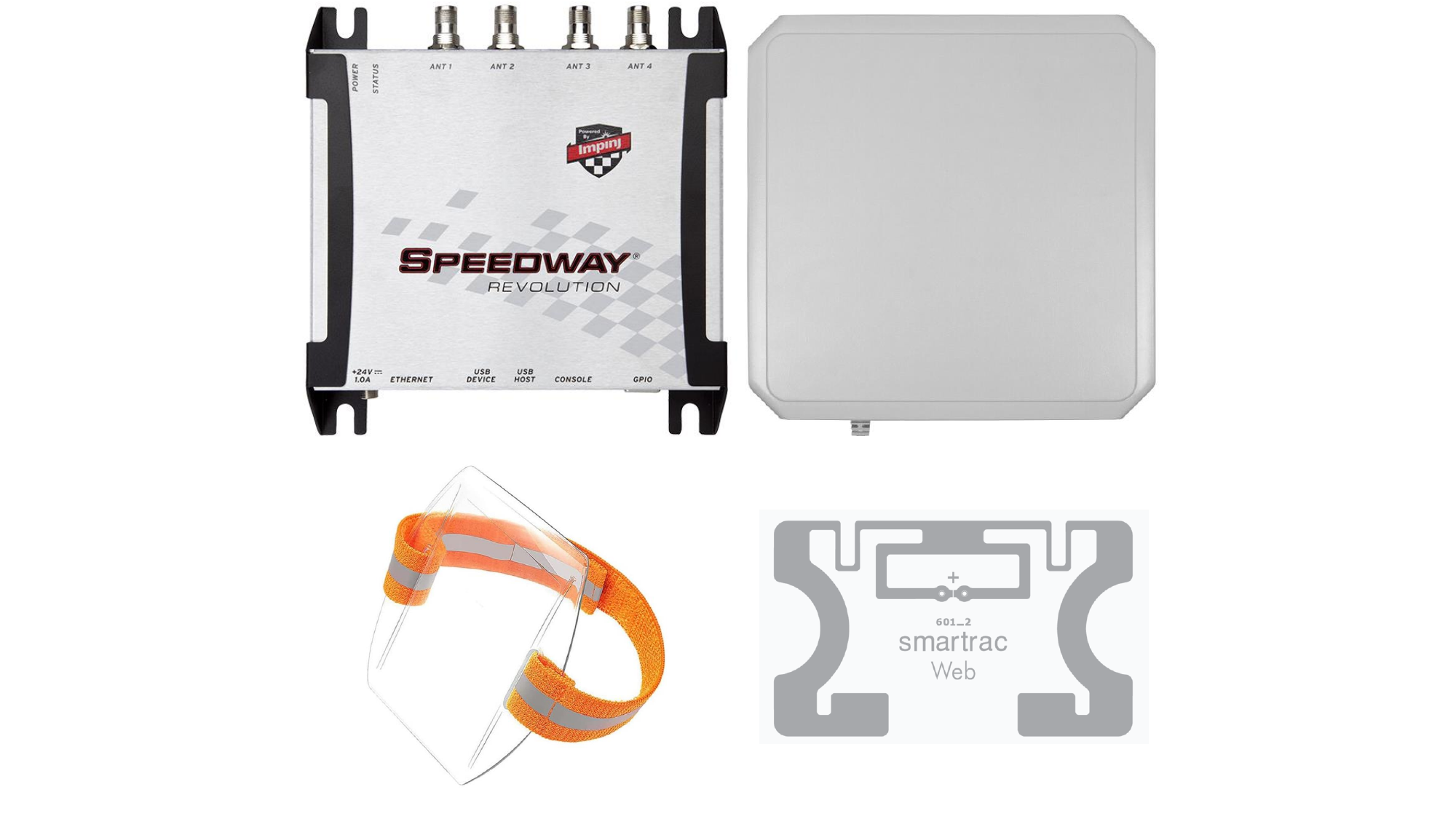}
	\caption{\small Hardware components of the real-time gait monitoring system used in deployment, including an RFID reader, two antennas, and a passive UHF tag in an armband. }
	\label{fig:hwconf}
	\vspace{-1.5em}
\end{figure}

The system supports registration of an arbitrary number of RFID tags via a configuration file that maps human-readable identifiers (e.g., `Tag1', `Tag2') to their 24-character EPC codes. In the current deployment, five tags are registered to support routine clinical operation and to allow replacement in cases of tag loss or damage. During measurement sessions, medical assistants select an available tag from the pool, secure it on the patient, and the system automatically initiates tracking when the tag enters the first antenna’s detection zone. Fig.~\ref{fig:hwconf} summarizes the hardware components used in deployment.

\subsection{Software Architecture}\label{subsubsec:software} 	

\begin{figure*}[t]
	\vspace{-2em}
	\centering
	\includegraphics[width = 1 \linewidth]{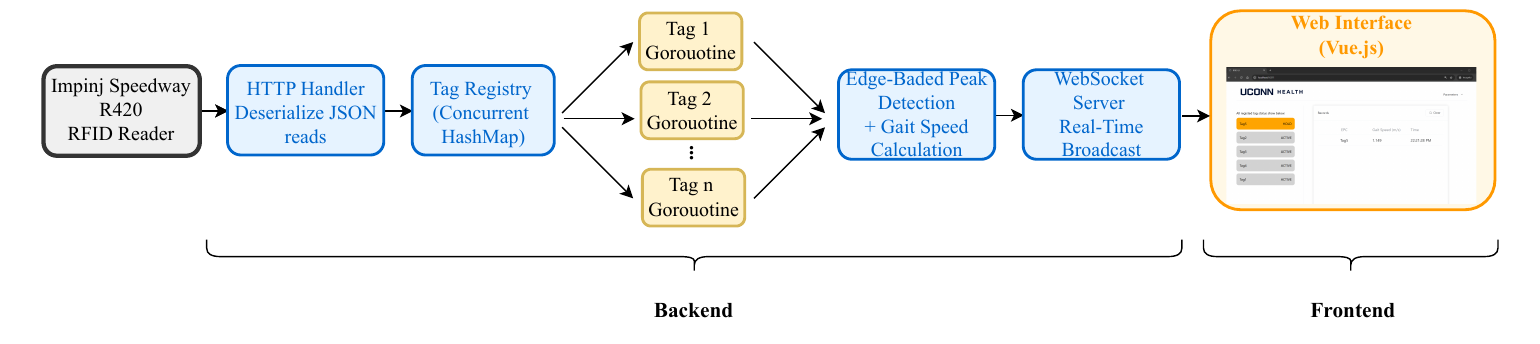}
	\caption{\small Software architecture and data flow for the real-time gait-speed monitoring system.}
	\label{fig:softarch}
	\vspace{-1em}
\end{figure*}

The software comprises a backend server that interfaces with the RFID reader, processes RSSI streams, and delivers computed results to a frontend web-based user interface (Fig.~\ref{fig:softarch}). The architecture prioritizes real-time processing, concurrent multi-patient tracking, and practical deployment in clinical environments.

\vspace{0.05in}

\noindent \textbf{Design Philosophy and Technology Selection.}
The system architecture reflects three guiding principles derived from clinical deployment requirements: universal accessibility, deployment simplicity, and real-time responsiveness.

The user interface (UI) is implemented as a web application using the Vue.js JavaScript framework and standard web technologies (HTML5, CSS, and JavaScript). This design addresses a fundamental challenge in clinical environments: the diversity of computing devices used by healthcare staff. A web-based UI runs in any modern browser across tablets, laptops, and desktop workstations regardless of operating system, eliminating software installation requirements and associated IT support burden. Clinical staff can access the gait monitoring system from bedside tablets during rounds or from nursing-station computers without configuration changes, supporting the workflow flexibility essential in healthcare settings.

The backend server is implemented in Go (Golang), selected for three characteristics particularly relevant to this application. First, Go provides lightweight concurrency primitives (goroutines and channels) that simplify the implementation of parallel processing pipelines, which naturally fit high-rate RFID read streams where multiple patients may be tracked simultaneously. Second, Go compiles to a single native binary without runtime dependencies, enabling straightforward deployment across Windows, Linux, or macOS clinical workstations via simple file distribution rather than complex installation procedures. Third, Go’s standard library provides robust support for network programming, including HTTP servers and WebSocket connections, reducing development complexity and external dependencies for the communication patterns this system requires.

\vspace{0.05in}
\noindent \textbf{Communication Architecture.}
The system employs two complementary communication protocols between the frontend and backend, each selected for its suitability to specific interaction patterns. REST API endpoints handle command-and-control operations, including patient registration, system configuration, and historical data retrieval. These request-response interactions align naturally with REST semantics, where the client initiates discrete operations and awaits confirmation. REST’s stateless design simplifies error handling and enables straightforward integration with clinical information systems should future interoperability be required.

WebSocket connections provide bidirectional, persistent channels for real-time data streaming. When a tag completes gait-speed computation, the result is immediately pushed to all connected user interfaces without polling overhead. This push-based communication ensures that clinical staff observe measurements with sub-second latency, supporting time-sensitive clinical workflows. Compared with repeated HTTP requests, the WebSocket protocol reduces network utilization and server load during continuous monitoring periods.

\vspace{0.05in}
\noindent \textbf{Concurrent Processing Architecture.} The backend adopts a goroutine-per-tag concurrency model in which each registered RFID tag maintains an independent processing context. A central tag registry, implemented as a concurrent hash map, maintains mappings from 24-character EPC identifiers to tag-handler instances. When RFID reads arrive from the reader, the system routes each read to the corresponding tag’s processing goroutine through a dedicated buffered channel.

This architecture provides three properties essential for clinical deployment. \textit{First}, isolation ensures that processing delays or errors affecting one tag do not impact concurrent measurements from other patients, a critical reliability requirement when multiple individuals are assessed simultaneously. \textit{Second}, the lightweight nature of Go goroutines (requiring only a few kilobytes of memory each) allows the system to support dozens of concurrent tags with minimal resource overhead. \textit{Third}, channel-based communication avoids explicit shared-state synchronization, reducing implementation complexity and the risk of concurrency errors.

\vspace{0.05in}
\noindent \textbf{Data Flow.}
The RFID reader transmits batches of tag reads to the backend via HTTP POST at a rate determined by its internal buffering (typically 5-15 batches per second during active reading). Each read record includes the tag’s EPC identifier, antenna port number, microsecond-precision timestamp, and RSSI value, encoded in JSON format. The HTTP handler deserializes incoming batches and routes individual reads to the appropriate tag goroutines.

Each tag channel maintains a buffer capacity of 1{,}024 reads, accommodating temporary processing delays without blocking the HTTP handler. This buffering prevents slow processing for one tag from delaying HTTP responses to the reader, which could otherwise cause timeouts or dropped batches. During typical operation, a single walking trial generates approximately 150-250 individual reads distributed across 10-20 HTTP batches. Upon completion of gait-speed computation, results are immediately pushed to connected interfaces via WebSocket, providing real-time feedback to clinical staff.

\section{Dual-Antenna Peak Detection Algorithm}
\label{sec:algorithm}
The core algorithmic challenge in dual-antenna gait-speed measurement stems from the live streaming nature of the RSSI feed and the spatial separation of antennas. Unlike offline signal processing, where the full signal trace is available in advance, our system processes RSSI readings as they arrive from the RFID reader without prior knowledge of when a walking event will begin or when the stream will terminate.

This section presents our solution from observation to implementation. We begin by characterizing real-world RFID signal behavior through offline analysis of deployment data (Section~\ref{subsubsec:signalchar}), revealing three critical properties: distance-dependent RSSI trends, plateau regions near peak signal strength, and noise fluctuations in clinical environments. These observations motivate our key algorithmic design principle, which exploits antenna beam symmetry to enable edge-based timing and avoid the ambiguity of plateau midpoint detection.  Section~\ref{subsubsec:designchalleng} identifies two core system challenges: the presence of multiple local maxima in the first antenna’s RSSI trajectory and the lack of a known signal termination point in real-time streaming. Section~\ref{subsubsec:movingwindow} presents our moving-window peak-detection algorithm, which realizes edge-based timing through asymmetric processing. Specifically, reverse-order detection is applied to the first antenna, while forward-order detection is used for the second antenna. Window sizes and detection thresholds are carefully selected to balance robustness against noise with timing precision.

\begin{figure}[t]
	\vspace{-0.6em}
	\centering
	\includegraphics[width = 1 \linewidth]{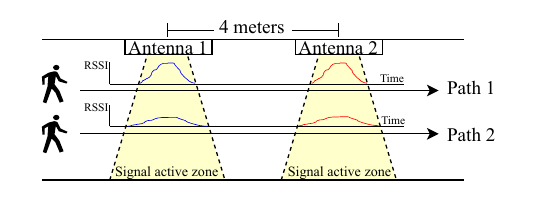}
	\caption{\small RSSI trajectories for two walking paths past an antenna, illustrating the distance-dependent rise-and-fall signal pattern.  }
	\label{fig:twopaths}
	\vspace{-1em}
\end{figure}

\subsection{Signal Streaming Characteristics}\label{subsubsec:signalchar}

To understand RFID signal behavior under real deployment conditions and guide algorithm design, we collected RSSI data from experimental walks performed by the system designer at the clinical deployment sites using the deployed hardware configuration. Prior to deployment, we conducted basic laboratory tests to validate reader configuration and tag readability. We focus our analysis on data collected directly in the clinical environment because RFID signal behavior is strongly environment-dependent. In particular, multipath reflections, RF interference, and physical layout characteristics can differ substantially between controlled laboratory settings and real clinical environments. This offline analysis allows us to examine complete signal sequences without real-time streaming constraints and to establish ground-truth timing for peak detection.

\vspace{0.05in}
\noindent 
\textbf{General signal trend.} Analysis of recorded signals shows a consistent distance-dependent pattern: as a tag approaches an antenna, the reported RSSI increases; after the tag passes the antenna, the RSSI decreases. Because RSSI is strongest near the antenna, this rise-and-fall pattern provides a reliable basis for timing-based gait-speed estimation. Fig.~\ref{fig:twopaths} illustrates this trend using a conceptual representation of RSSI trajectories for two walking paths.

\begin{figure}[t]
	\vspace{0.7em}
	\centering
	\includegraphics[width = 0.96 \linewidth]{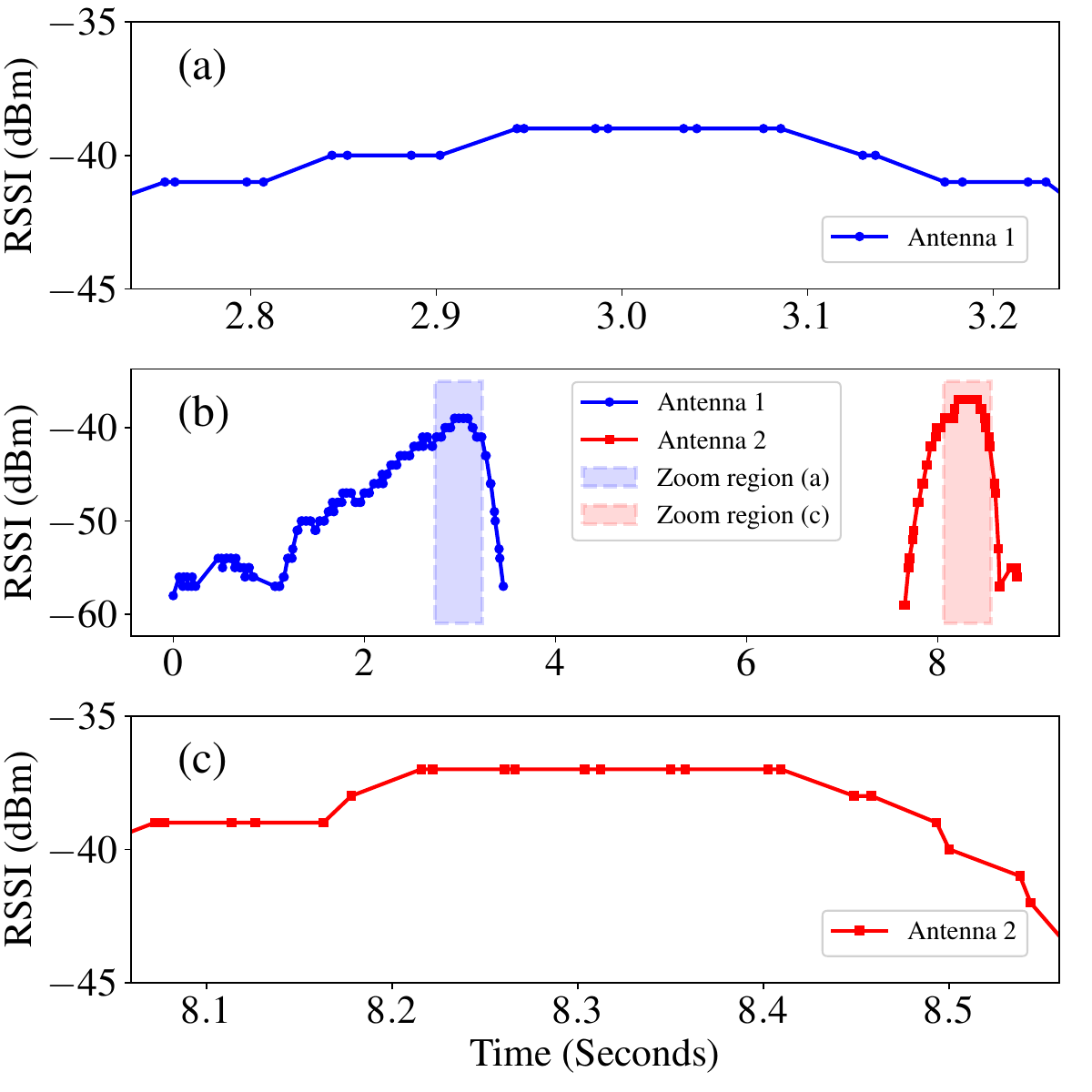}
	\caption{\small RSSI traces from both antennas during a normal walk, with zoomed-in views of the plateau regions near the peaks. }
	\label{fig:highlightedwindow}
\end{figure}

\vspace{0.05in}
\noindent \textbf{Plateau behavior at peak.}
An important observation from the recorded signals is the presence of plateau regions in which the RSSI remains relatively constant near its maximum value. Rather than exhibiting a sharp, well-defined peak, the RSSI often forms a flat region around the maximum. Fig.~\ref{fig:highlightedwindow}(b) shows representative RSSI traces from both antennas during a walking trial, while Fig.~\ref{fig:highlightedwindow}(a) and Fig.~\ref{fig:highlightedwindow}(c) provide zoomed-in views of the highlighted peak regions. The plateau for Antenna~1 spans eight consecutive samples, while the plateau for Antenna~2 spans ten consecutive samples.

The plateau arises because RSSI changes rapidly with position when the tag is far from the antenna but varies only slightly when the tag passes very close to it. As the subject traverses this near-antenna region, small changes in tag-to-antenna distance produce minimal RSSI variation, causing multiple consecutive samples to report nearly identical values. This physical effect yields a plateau near the peak of the RSSI trajectory rather than a sharp, well-defined maximum. Transmit power influences the length of this plateau: reducing transmit power can shorten the plateau by shrinking the saturation region near the antenna, whereas increasing power can further flatten the peak. However, transmit power cannot be arbitrarily reduced, as insufficient power may prevent reliable tag detection. In practice, power settings must balance coverage and stability, and plateau behavior persists under typical operating conditions.

These plateau regions directly affect how peak timing can be estimated from RSSI. A natural approach is to use the midpoint of the plateau as an approximation of the peak location. However, the plateau length varies across walks and cannot be determined in advance during real-time streaming, making midpoint-based estimation unreliable.

\begin{figure}[t]
	\vspace{-0.4em}
	\centering
	\includegraphics[width = 1 \linewidth]{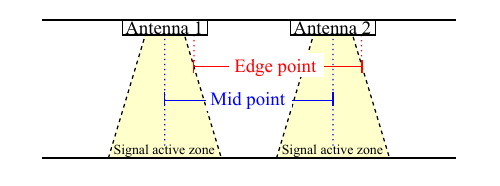}
	\caption{\small Comparison of midpoint-based and edge-based trigger timing.}
	\label{fig:edgesymmetry}
\end{figure}

\vspace{0.05in}
\noindent \textbf{Antenna beam symmetry and edge-based timing.}
Given the challenges of identifying a single peak moment within plateau regions and measurement noise that can mask small signal changes, we exploit a fundamental property of the antenna radiation pattern: beam symmetry. Because both antennas are identical models with identical mounting orientations and RF settings, their radiation patterns are effectively symmetric. To motivate our approach, Fig.~\ref{fig:edgesymmetry} contrasts midpoint-based timing (which depends on observing the full plateau) with edge-based timing (which can be triggered online).

The key insight is that this symmetry allows us to use edge-based timing instead of locating the ambiguous peak midpoint. Because both antennas share the same radiation pattern, the temporal offset between the peak midpoint and the descending edge is consistent across antennas. We therefore detect the right edge of each antenna’s RSSI peak, defined as the point at which the signal has clearly descended from the plateau region. As a result, the time difference between the midpoints of the two antenna peaks equals the time difference between their right edges.

\subsection{Algorithm Design Challenges}\label{subsubsec:designchalleng}
The signal characteristics described above introduce specific design challenges for real-time processing of RSSI streams. In particular, the algorithm must (i) select the correct peak when multiple local maxima occur and (ii) detect peak timing without prior knowledge of when a walking event and its RSSI stream will end.

\begin{figure}[t]
	\centering
	\includegraphics[width = 0.96 \linewidth]{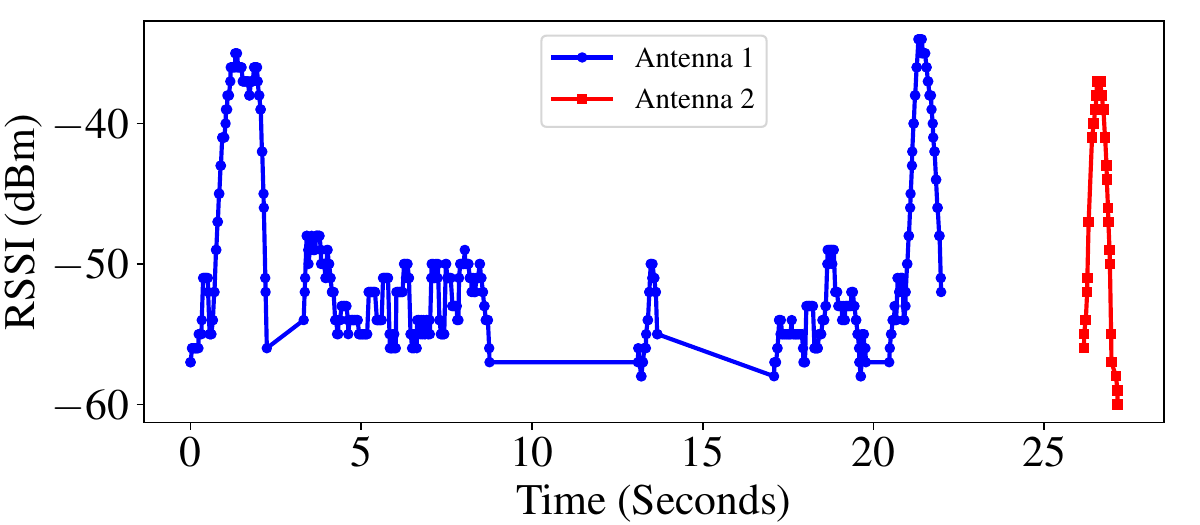}
	\caption{\small A recording example on multiple peaks.  }
	\label{fig:multiplepeaks}
	\vspace{-0.5em}
\end{figure}

\vspace{0.05in}
\noindent \textbf{Multiple peaks at first antenna.}
When a patient enters the measurement zone, the first antenna may detect the tag multiple times before the patient reaches the optimal detection range. Unlike sensing systems with sharply defined boundaries---such as camera-based systems with fixed fields of view or pressure mats with explicit contact regions---an RFID antenna’s active zone does not form a clear spatial boundary. As a result, a tag may unintentionally enter and exit the active region before the actual walking event, and all such signal readings are continuously forwarded to the backend. Fig.~\ref{fig:multiplepeaks} shows an example in which the recorded RSSI trace contains multiple peaks before and during a single walking pass. Although several peaks are observed, only the final peak from Antenna~1 and the first peak from Antenna~2 correspond to the intended gait measurement event.

\vspace{0.05in}
\noindent 
\textbf{Unknown signal termination.}
In real-time streaming operation, the system cannot determine when the signal from the first antenna will end because the RFID reader continuously reports tag reads and provides no explicit marker for the completion of a walking trial. A patient may enter the measurement zone, pause, turn around, or restart the attempt, resulting in RSSI observations at the first antenna that persist for an unknown duration.

To resolve this uncertainty, we leverage the spatial separation of the antennas. Each RFID read includes the active antenna port, a timestamp, and the RSSI value. Using the antenna port information, the system distinguishes which antenna produced each reading and uses the first detection at the second antenna as a completion trigger for processing the accumulated data from the first antenna. At this point, the first antenna’s RSSI readings form a complete signal sequence, enabling reliable identification of the true peak despite multiple local maxima. The second antenna thus serves a dual purpose: providing the exit timing measurement and triggering event-driven processing of the first antenna’s data.

\subsection{Moving Window Peak Detection}\label{subsubsec:movingwindow}
Having established that edge-based timing provides equivalent accuracy to midpoint detection while avoiding plateau ambiguity, we now describe our moving window algorithm that detects the right edge of each antenna's RSSI peak under real-time streaming constraints.

\vspace{0.05in}
\noindent \textbf{First antenna detection (reverse-order processing).}
The first antenna processes RSSI readings in reverse chronological order to address the multiple-peak challenge described above. Intuitively, the true walking pass corresponds to the last strong peak observed at Antenna~1 before the tag is first detected by Antenna~2; reverse-order processing encounters this peak first and avoids earlier false peaks.

When the second antenna detects the tag---signaling that the patient has passed through the first antenna’s measurement zone---the algorithm reverses the complete sequence of RSSI readings accumulated from the first antenna and applies a moving window of size $W_1$, processing from the most recent sample backward toward the oldest.

At each position in the reversed sequence, the algorithm maintains a window of $W_1$ samples (in reverse chronological order). Within this window, it identifies the maximum RSSI value $r_{\max}$ and compares it with the RSSI of the current sample at the window edge, $r_{\text{current}}$. When the difference exceeds a threshold $\tau_1$, indicating that the signal has descended from the peak region, the algorithm records the timestamp $t_{\text{start}}$ corresponding to $r_{\max}$ within that window. This timestamp represents the right edge of the first antenna’s peak—the last moment before the descending phase begins. Pseudocode for right-edge detection at the first antenna is provided in Alg.~\ref{alg:firstantenna}.

\begin{algorithm}[t]
\caption{First Antenna Right Edge Detection}
\label{alg:firstantenna}
\KwIn{Reversed RSSI sequence $R[1..N]$, window size $W_1$, threshold $\tau_1$}
\KwOut{Start timestamp $t_{\text{start}}$}

\tcp{Process in reverse chronological order}
Reverse data: $R[i] \gets S[N-i+1]$ for $i = 1$ to $N$\;

$W \gets \emptyset$\;

$t_{\text{start}} \gets \texttt{NULL}$\;

\For{$i \gets 1$ \KwTo $N$}{
    Append $(R[i].\text{rssi},\, R[i].\text{timestamp})$ to window $W$\;
    
    \If{$|W| > W_1$}{Remove oldest sample from $W$}

    \If{$|W| = W_1$}{
    \tcp{Initialize with first sample in window}
    $r_{max} \gets W[0].\text{rssi}$, $t_{max} \gets W[0].\text{timestamp}$\;

    \ForEach{sample $s \in W[1..|W|]$}{
            \If{$s.\text{rssi} > r_{max}$}{
                $r_{max} \gets s.\text{rssi}$, $t_{max} \gets s.\text{timestamp}$\;
            }
        }
    
    
    $r_{current} \gets W[\text{last}].\text{rssi}$\;

    \If{$r_{max} - r_{current} \geq \tau_1$}{
        \Return{$t_{max}$}\tcp*{Return timestamp of peak RSSI}
    }
    }
}
\Return{$t_{start}$}\;
\end{algorithm}

By processing in reverse order, the algorithm encounters the true peak (which occurs last in time) before any earlier false peaks, ensuring reliable detection of the correct starting time. Once the drop condition is satisfied, the algorithm terminates immediately and ignores all remaining earlier samples.

\vspace{0.05in}
\noindent 
\textbf{Second antenna detection (forward-order processing).}
The second antenna processes RSSI readings in forward chronological order as they arrive in real time. Because the second antenna only detects tags after they have passed the first antenna, multiple early peaks are not a concern. During a normal walk, the patient is moving away from the first antenna and approaching the second. The algorithm applies the same moving-window approach with window size $W_2$ and threshold $\tau_2$, but processes samples in arrival order.

A key implementation detail distinguishes the second-antenna algorithm from the first: when the condition $r_{\max} - r_{\text{current}} \geq \tau_2$ is satisfied, the algorithm outputs $t_{\text{end}}$ using the timestamp of the second-to-last sample in the window (i.e., $t_{\text{end}} = T[i-1]$) rather than the current triggering sample (i.e., $T[i]$). This avoids a one-sample delay, because the descent is only confirmed at the triggering sample, whereas the right edge is defined as the last amoment before the descent begins.

\begin{figure}[t]
	\vspace{-0.8em}
	\centering
	\includegraphics[width = 0.96 \linewidth]{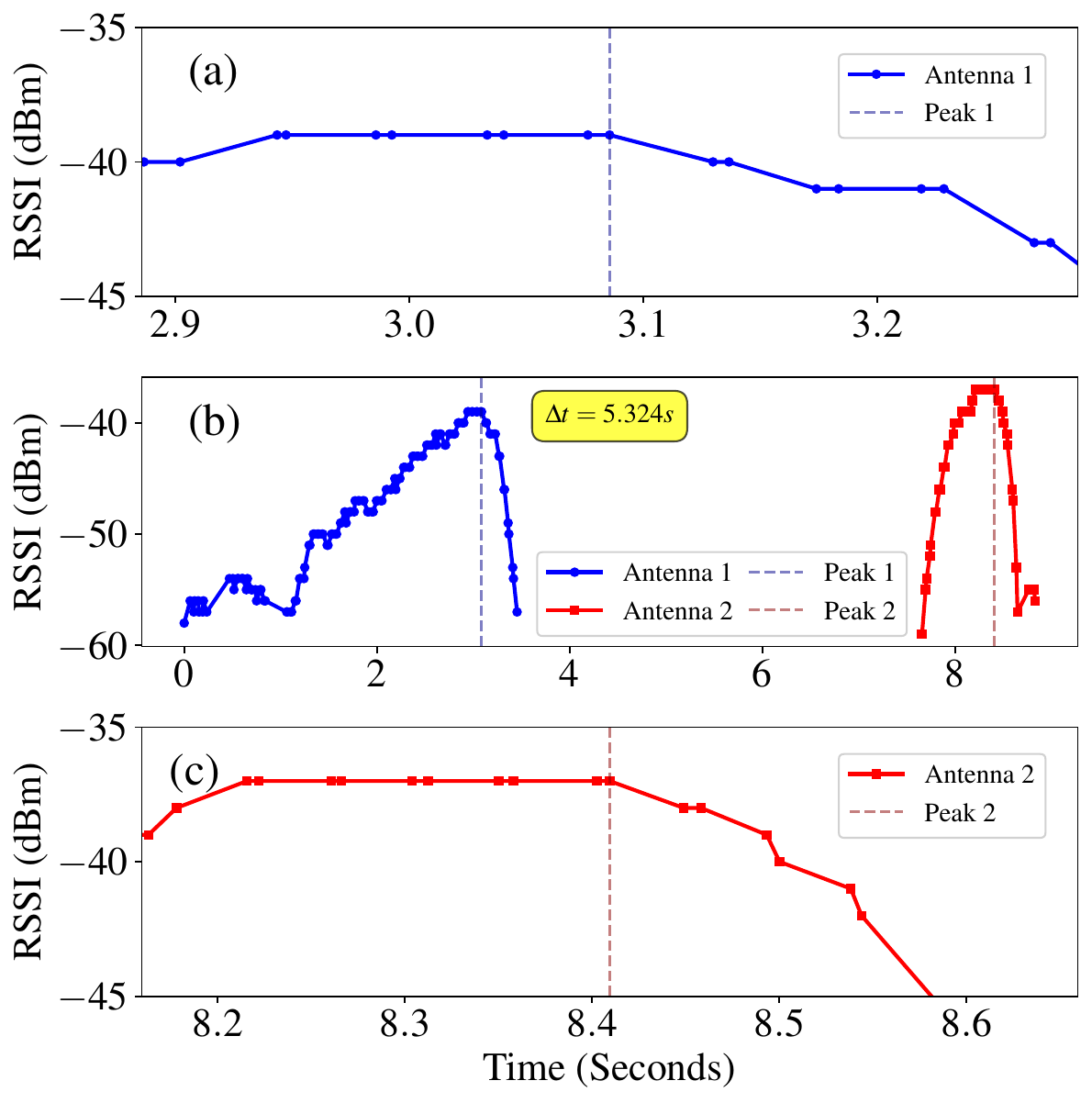}
	\caption{\small Demonstration of the edge-based peak detection algorithm on an example walk.}
	\label{fig:demonstration}
	\vspace{-1em}
\end{figure}

\vspace{0.05in}
\noindent 
\textbf{Algorithm demonstration.} The complete edge-based peak-detection process for a representative walking trial is illustrated in Fig.~\ref{fig:demonstration}. Fig.~\ref{fig:demonstration}(b) shows the full RSSI traces from both antennas as a subject traverses the 4~m measurement zone. The blue curve corresponds to Antenna~1 and captures the approach to and passage through the first antenna, while the red curve corresponds to Antenna~2 and captures the subsequent passage through the second antenna. Vertical dashed lines mark the detected right edges of the RSSI peaks at $t_{\text{start}}$ and $t_{\text{end}}$, respectively.

Zoomed views of the peak regions in Fig.~\ref{fig:demonstration}(a) and (c) reveal the plateau behavior at both antennas. For Antenna~1, the RSSI remains nearly constant for approximately 0.2~s (8–10 samples at the current read rate), with less than 1~dBm variation across the plateau, followed by a clear and continuous descent identified at $t_{\text{start}} = 3.086$~s. A similar pattern is observed for Antenna~2, with the right edge detected at $t_{\text{end}} = 8.409$~s. The resulting time interval is $t_{\text{end}} - t_{\text{start}} = 5.324$~s and gait speed $v = 4\ m/5.324\ s = 0.751\ m/s$. This example demonstrates the algorithm’s ability to reliably extract precise timing from streaming RSSI data despite extended plateau regions and noise fluctuations.

\begin{figure}[t]
	\centering
	\includegraphics[width = 0.96 \linewidth]{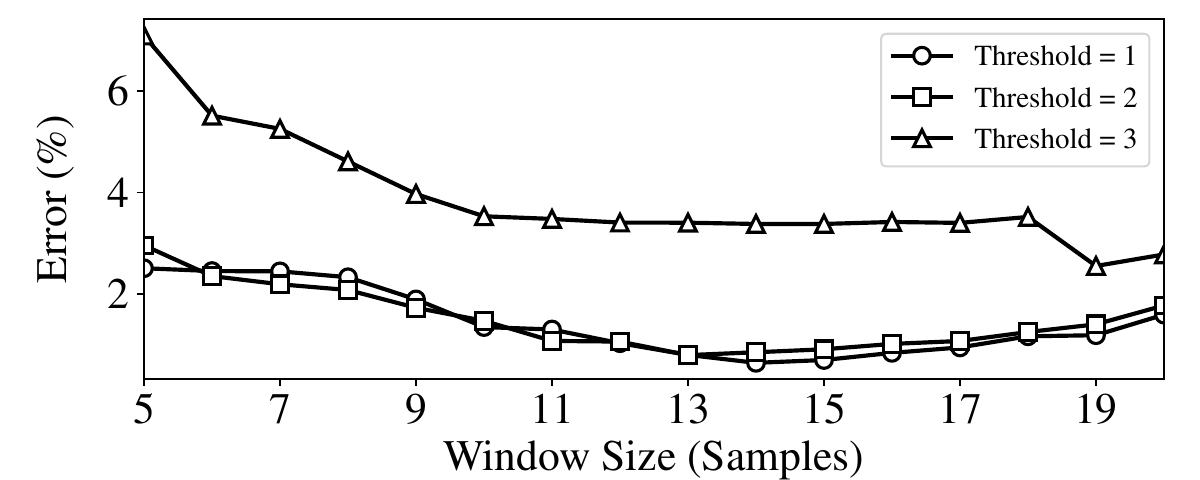}
	\caption{\small Effect of window size $W$ and drop threshold $\tau$ on gait-speed measurement error. }
	\label{fig:windowthreshold}
\end{figure}

\vspace{0.05in}
\noindent \textbf{Window size and threshold selection.} Based on the plateau behavior observed in Fig.~\ref{fig:demonstration}, the algorithm uses a moving window to span the plateau region and a drop threshold to robustly distinguish true signal descent from noise. To select these parameters, we recorded RSSI traces from repeated walks performed by the system developer in the same clinical hallway and using the same procedure as patient measurements (identical tag placement, antenna layout, and reader configuration). We first processed each recorded trace offline. Because the full RSSI sequence is available in this setting, we can locate peak edges without unknown-termination constraints and treat the resulting timestamps as an \emph{offline reference}. We then replayed the same recorded reads to the backend server in chronological order to emulate online operation. For each $(W,\tau)$ setting, we compared the server output with the offline reference to compute percentage error. In later clinical evaluation, we use concurrent stopwatch timing $v_{\mathrm{SW}}$ as the ground-truth baseline for gait-speed accuracy.

Fig.~\ref{fig:windowthreshold} shows the average gait-speed measurement error as a function of window size for three threshold values ($\tau = 1,2,3 $~dBm), computed across 26 walking trials performed by the system developer at the clinical deployment site. Measurement error is defined as: 
$\left|\frac{v_{\text{measured}} - v_{\text{ref}}}{v_{\text{ref}}}\right| \times 100\%$, where $v_{\text{ref}}$ denotes the offline reference gait speed.

Three key observations emerge from this analysis:
\begin{enumerate}
    \item \textbf{Effect of window size.}
    Error decreases sharply as the window size increases from $W=5$ (2.5\% error for $\tau=1$) to $W=14$ (0.65\% error). Beyond $W=14$, larger windows provide little additional improvement and may slightly increase error. Small windows ($W<10$) often fail to span the full plateau region (typically 8--12 samples under our reader configuration), so noise within the plateau can be mistaken for a true signal drop, causing the algorithm to trigger too early.

    \item \textbf{Threshold sensitivity.}
    The drop threshold $\tau$ has a significant impact on accuracy. $\tau=1$~dBm yields the lowest error across all window sizes (0.65--2.5\%), while $\tau=3$~dBm consistently produces higher error (3.5--7\%) due to delayed edge detection that shifts the detected timing beyond the true peak boundary. $\tau=2$~dBm provides intermediate performance but offers no advantage over $\tau=1$~dBm.

    \item \textbf{Parameter robustness.}
    Error remains below 1.5\% across a broad range of settings ($W=10$--17, $\tau=1$--2~dBm), indicating that the algorithm does not require precise tuning to achieve reliable performance.
\end{enumerate}

Based on these results, we adopt $W_1=W_2=14$ and $\tau_1=\tau_2=1$~dBm for on-site deployments.

\noindent \textbf{Baseline Comparison}
To evaluate the effectiveness of the proposed edge-based peak-detection algorithm, we compare it against a threshold-crossing baseline. 
The baseline follows the same asymmetric processing structure as the proposed method, which is reverse-order scanning for the entry antenna and forward-order scanning for the exit antenna, but replaces the moving-window drop trigger with a fixed RSSI threshold. 

To identify optimal operating points, we perform a exhaustive search over 31 threshold values ranging from $-70~dBm$ to $-40$~dBm in 1~dBm step size. Table~\ref{tab:baseline_comparison} reports these best three threshold settings alongside three representative configurations of the proposed method. Even at its best operating points, the threshold-crossing baseline produces substantially larger errors, with MAE values ranging from 0.1006 to 0.1082 and and error rates of 10.94\%–11.30\%. In addition, the baseline fails to produce valid measurements consistently, achieving success rates of only 58\%, 65\%, and 81\%, corresponding to 5–11 missing outputs across the 26 trials. Fixed RSSI thresholds are prone to missed detections and cannot robustly capture the distance-dependent signal behavior that RSSI increases as the tag approaches the antenna and decreases as it moves away. In contrast, the proposed method achieves 100\% success with significantly lower error by detecting timing events at stable RSSI peak edges via a sliding-window drop, making it robust to signal fluctuations and eliminating the need for threshold tuning.

\begin{table}[t]
\vspace{0.7em}
\centering
\caption{Comparison of the proposed peak-detection method against the
threshold-crossing baseline.}
\label{tab:baseline_comparison}
\small
\setlength{\tabcolsep}{4pt}
\renewcommand{\arraystretch}{1.15}
\begin{tabular}{llccc}
\toprule
  \textit{Method} & \textit{Setting} & \textit{\shortstack{MAE \\ (m/s)}} & \textit{\shortstack{Error \\ Rate (\%)}} & \textit{\shortstack{Success \\ Rate}} \\
\midrule
\multicolumn{5}{c}{\textbf{Threshold Baseline}} \\
\midrule
  Baseline \#1 & RSSI $\geq$ -40 dBm & 0.1055 & 10.94 & 15/26 (58\%) \\
  Baseline \#2 & RSSI $\geq$ -45 dBm & 0.1006 & 11.02 & 21/26 (81\%) \\
  Baseline \#3 & RSSI $\geq$ -41 dBm & 0.1082 & 11.30 & 17/26 (65\%) \\
\midrule
\multicolumn{5}{c}{\textbf{Proposed Method }} \\
\midrule
  Proposed \#1 &$W=14$, $\tau=1$ & 0.0060 & 0.65 & 26/26 (100\%) \\  
  Proposed \#2 &$W=15$, $\tau=1$ & 0.0066 & 0.70 & 26/26 (100\%) \\
  Proposed \#3 &$W=13$, $\tau=2$ & 0.0075 & 0.80 & 26/26 (100\%) \\
\bottomrule
\end{tabular}
\end{table}

\section{Clinical Deployment and Evaluation}
\label{sec:deployment}

\subsection{Clinical Deployment}

\begin{figure*}[t!]
	\centering
	\subfloat[\label{fig:WFgeri}]{
		\includegraphics[width=0.18\textwidth, height=2.44cm, keepaspectratio]{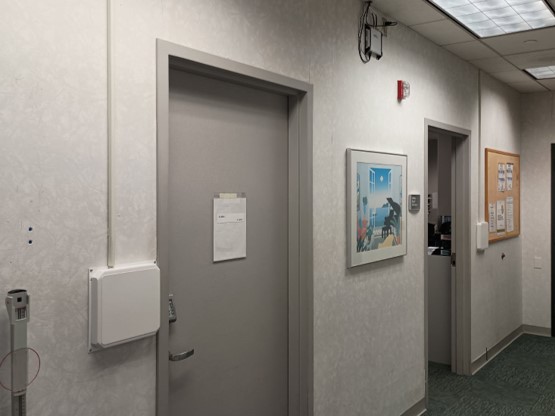}
	}
	\hspace{-0.1em}
	\subfloat[\label{fig:WFReynolda}]{
		\includegraphics[width=0.18\textwidth]{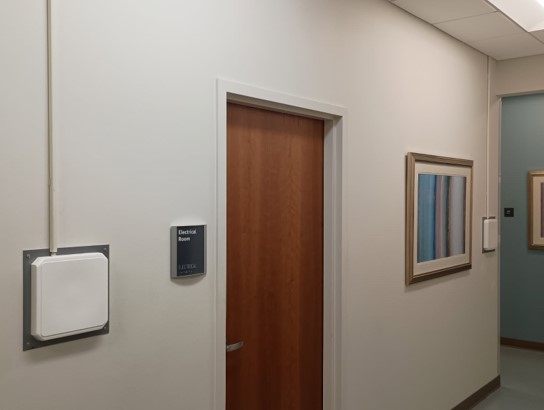}
	}
	\hspace{-0.1em}
	\subfloat[\label{fig:UConnSouth}]{
		\includegraphics[width=0.18\textwidth]{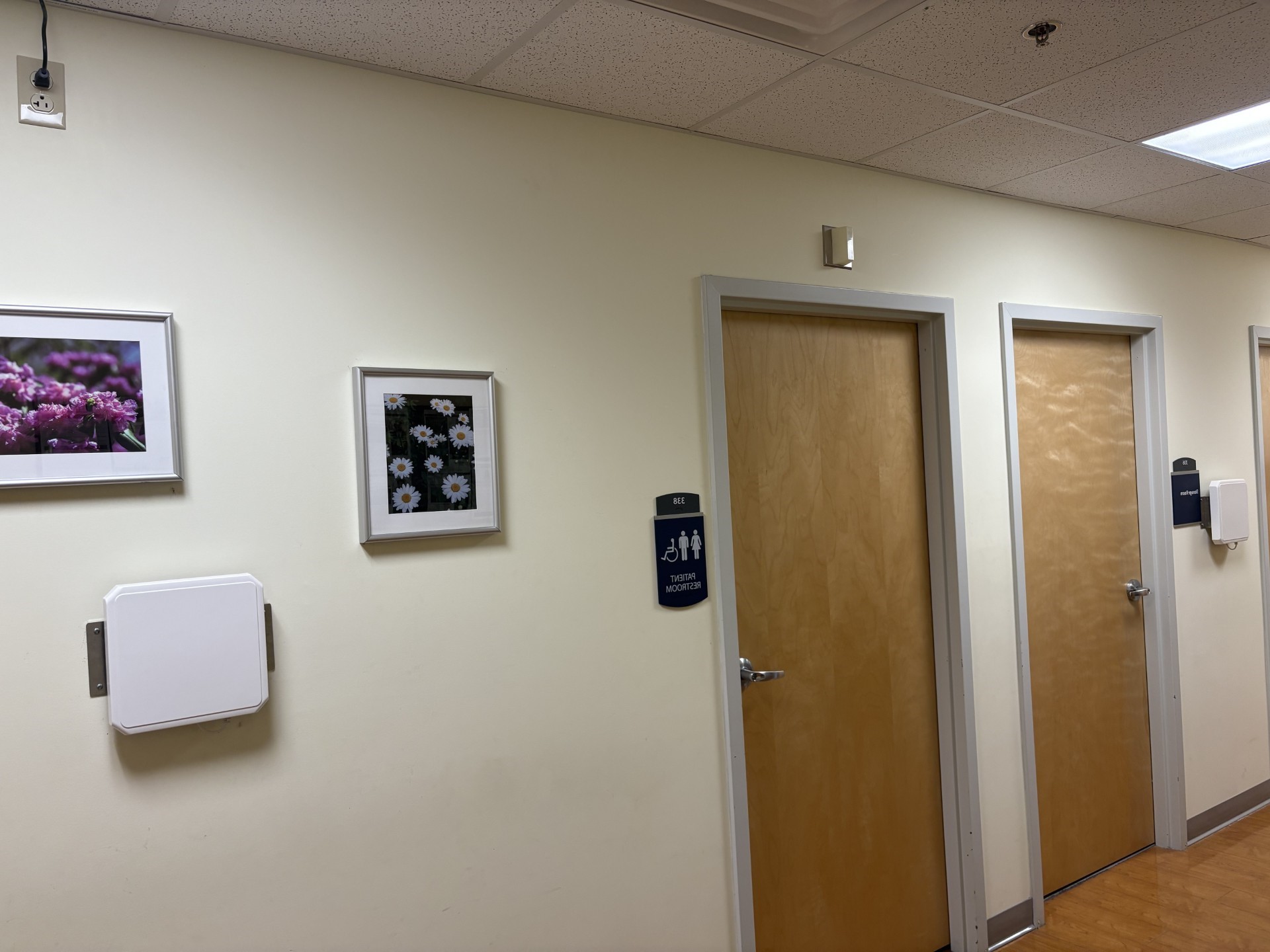}
	}
	\hspace{-0.1em}
	\subfloat[\label{fig:UConngeri}]{
		\includegraphics[width=0.18\textwidth]{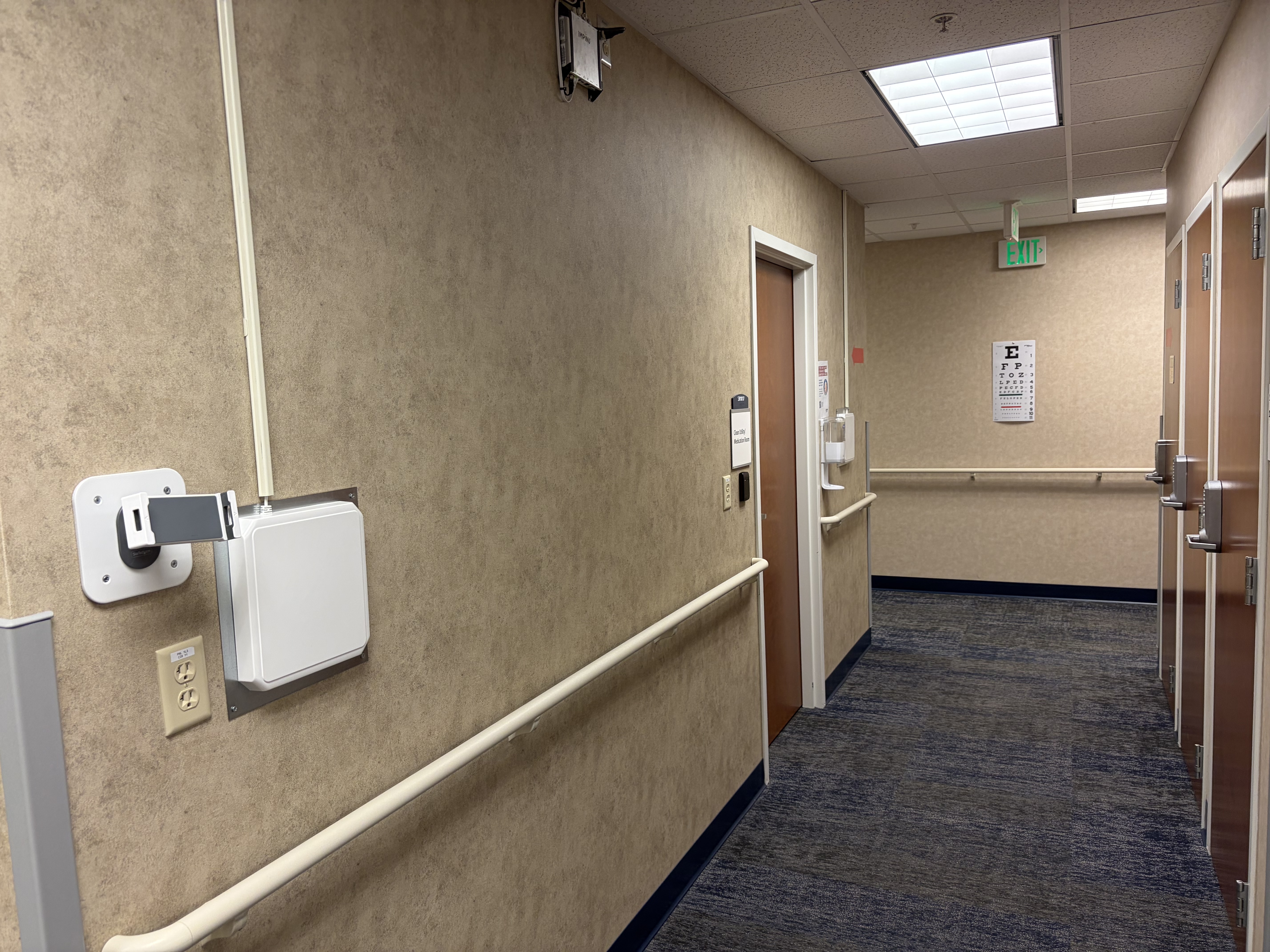}
	}
	\hspace{-0.1em}
	\subfloat[\label{fig:UConnOP}]{
		\includegraphics[width=0.18\textwidth]{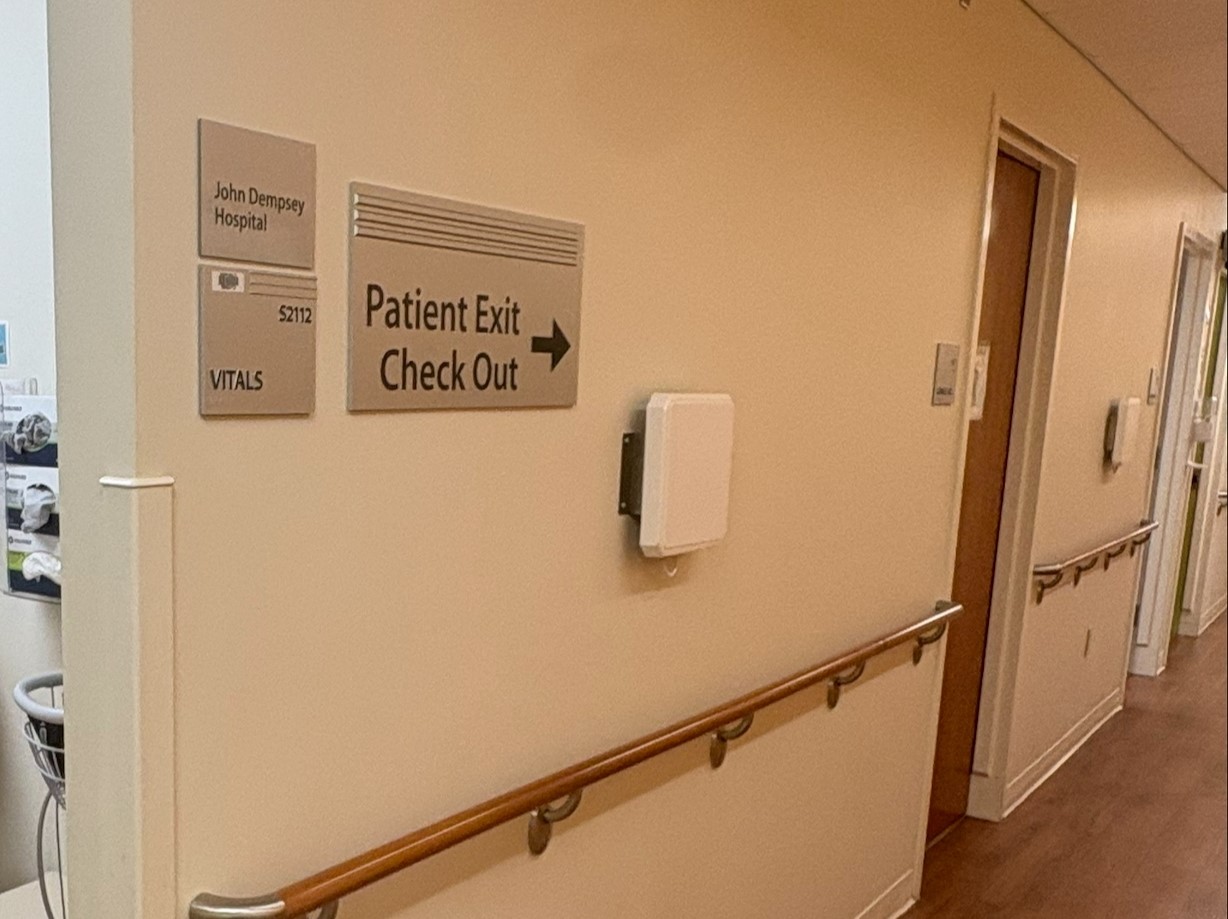}
	}
	\caption{\small RFID antenna placement and hallway configuration across deployment sites:
	(a)~Atrium Health Wake Forest Baptist Geriatric Medicine (Sticht Center),
	(b)~Atrium Health Wake Forest Baptist Family Medicine (Reynolda),
	(c)~UConn Health Internal Medicine (Southington),
	(d)~UConn Center for Healthy Aging and Geriatrics,
	(e)~UConn Health Infectious Diseases (Outpatient Pavilion).}
	\label{fig:deploymentphotots}
\end{figure*}

We deployed the RFID-based gait-speed monitoring system across three UConn Health outpatient clinics to evaluate performance under routine clinical operation: (i) UConn Center for Healthy Aging and Geriatrics, (ii) UConn Health Internal Medicine at Southington, and (iii) UConn Health Infectious Diseases at Outpatient Pavilion. In addition to UConn Health, the system was also installed at two Atrium Health Wake Forest Baptist clinics (Geriatric Medicine at the Sticht Center and Family Medicine at Reynolda). However, data from those sites are not included in the quantitative evaluation due to limited availability. Fig.~\ref{fig:deploymentphotots} shows the antenna placement and hallway configuration at each deployment site. 

Deploying the system in active clinical environments required addressing several practical constraints. At each site, antenna placement was determined by a straight, unobstructed hallway segment near the patient intake area that could support a dedicated walking path. Both antennas were mounted at a height of 1.2~$m$, which provides consistent line-of-sight coverage across varying patient heights. The RFID reader and backend PC were positioned at a nearby clinical workstation and connected to the antennas via coaxial cables routed along walls or hidden within the ceiling to eliminate tripping hazards. The 4~$m$ antenna spacing was selected to provide sufficient walking distance for clinically meaningful gait assessment~\cite{maggio2016instrumental}, while also fitting within the available hallway lengths across deployment sites.
This distance remains configurable in the system software to accommodate shorter corridors if necessary. 

Importantly, installation at each site was completed within a single session and required no modifications to existing clinical infrastructure, minimizing operational disruption and facilitating practical deployment. 

\begin{figure}[t]
	\centering
	\includegraphics[width = 0.96 \linewidth]{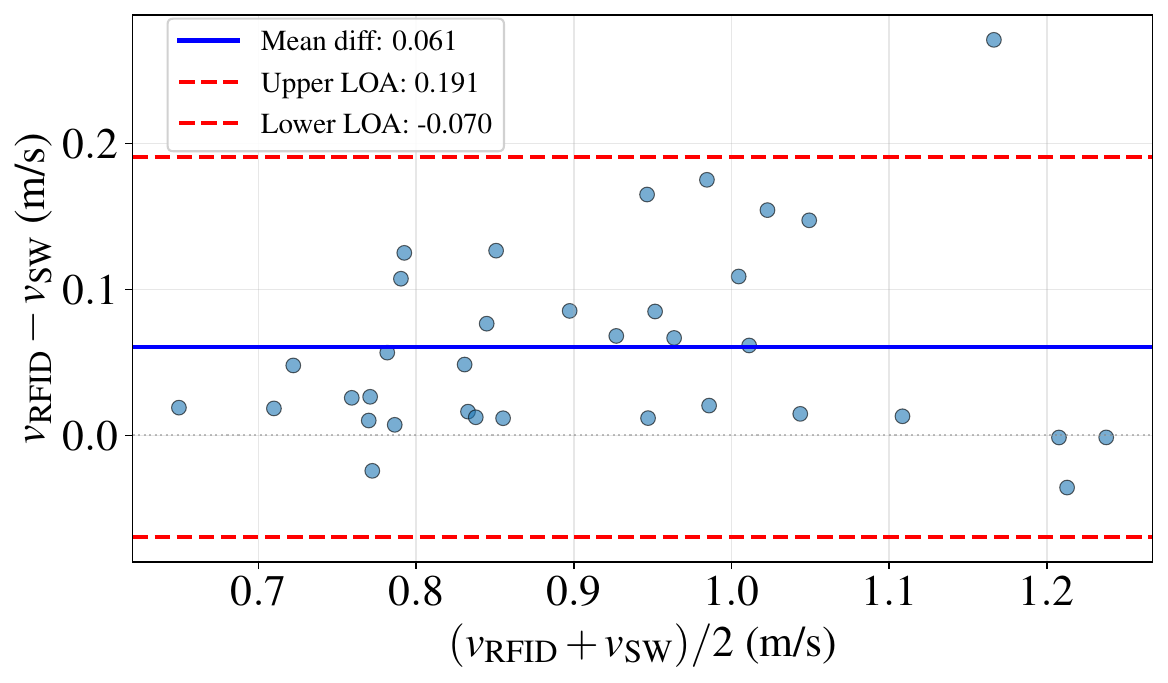}
	\caption{\small  Bland--Altman plot comparing RFID-based gait-speed estimates $v_{\mathrm{RFID}}$ with stopwatch-derived speeds $v_{\mathrm{SW}}$ across 35 walks.}
	\label{fig:bland}
\end{figure}

\subsection{Clinical Evaluation}

Data at the Geriatrics clinic were collected from February to July, while data from the other two clinics were collected from September to November, yielding 966 gait-speed trials in total. Across sites, the clinical workflow is consistent:  after completing standard vital sign measurements, a medical assistant secures a passive RFID tag in an armband on the patient’s upper arm, instructs the patient to walk past two antennas mounted 4~m apart in the hallway, and the system reports gait speed immediately in the browser-based interface. At the UConn Health Infectious Diseases clinic, gait-speed assessment is intended for all patients aged 50 and older, whereas in the other clinics measurements are typically collected during Medicare Annual Wellness Visits. Measurements are obtained during routine care rather than under a laboratory protocol and therefore reflect realistic clinical conditions rather than idealized experimental settings.

\begin{table}[t]
\vspace{0.5em}
\centering
\caption{Gait-speed measurement outcomes across three clinical deployment sites. }
\label{tab:clinic}
\small
\setlength{\tabcolsep}{2pt}
\renewcommand{\arraystretch}{1.15}
\begin{tabular}{lccccc}
\toprule
  & \textit{\shortstack{Trials \\ (n)}}
  & \textit{\shortstack{Successful \\ measurements}}
  & \textit{\shortstack{System \\ failure}}
  & \textit{\shortstack{Erroneous \\ measurements}}
  & \textit{\shortstack{Excluded \\ trials}} \\

\midrule
\multicolumn{6}{c}{\textbf{UConn Center for Healthy Aging and Geriatrics}} \\
\midrule
  & 361 & 312 (86.4\%) & 22 (6.1\%) & 25 (6.9\%) & 2 (0.6\%) \\
\midrule
\multicolumn{6}{c}{\textbf{UConn Health Internal Medicine (Southington)}} \\
\midrule
  & 156 & 148 (94.9\%) & 3 (1.9\%) & 5 (3.2\%) & 0 (0.0\%) \\
\midrule
\multicolumn{6}{c}{\textbf{UConn Health Infectious Diseases (Outpatient Pavilion)}} \\
\midrule
  & 449 & 387 (86.2\%) & 20 (4.5\%) & 13 (2.9\%) & 29 (6.5\%) \\
\bottomrule
Total  & 966 & 847 (87.7\%) & 45 (4.7\%) & 43 (4.5\%) & 31 (3.2\%)\\
\bottomrule
\end{tabular}
\end{table}
Because routine clinical workflows do not provide ground-truth timing, we first evaluate deployment performance using system success rate and data-quality outcomes, summarized in Table~\ref{tab:clinic}. \textit{Trials} denotes the number of walking attempts recorded at each clinic. \textit{Successful measurements} are trials that produced a meaningful gait-speed estimate. \textit{System failures} indicate cases in which the system did not produce a valid output (reported speed = 0). \textit{Erroneous measurements} are implausible gait-speed estimates outside the expected clinical range (speed $<0.2$~$m/s$ or $>2.0$~$m/s$). \textit{Excluded trials} are cases in which the system output was 0 but the medical assistant did not document a reason (e.g., the walk was not performed or the measurement was missed).  Overall, the system achieved an 87.7\% successful measurement rate across all clinics. As this was an early deployment, these results reflect initial real-world performance.  A known limitation is that shorter patients may position the tag below the effective antenna beam at the 1.2~$m$ mounting height,  resulting in insufficient signal strength for reliable peak detection.  Because the antenna beam widens with distance from the antenna, instructing shorter patients to walk farther from the antenna wall increases the likelihood that the tag falls within the active detection zone, providing a practical mitigation for this issue. We expect further improvements in system reliability and measurement quality as deployment procedures mature and the system is refined. 

To evaluate measurement accuracy, we conducted a separate comparison in which the system developer performed 35 walks at varying speeds at one of the clinical deployment sites, using the same hardware configuration, antenna layout, and tag placement as patient measurements.
Each trial was timed concurrently with a manual stopwatch, yielding paired speed estimates $v_{\mathrm{RFID}}$ (system output) and $v_{\mathrm{SW}}$ (stopwatch-derived speed, computed as $4~\mathrm{m}\,/\,t_{\mathrm{SW}}$). Across all trials, the system achieves a mean absolute error (MAE) of 0.064~$m/s$ compared with stopwatch timing.

To assess whether the level of agreement between the two methods depends on gait speed, we use a Bland--Altman analysis~\cite{bland2010statistical}, the standard approach for comparing two measurement instruments.  Fig.~\ref{fig:bland} plots the difference between methods $v_{\mathrm{RFID}} - v_{\mathrm{SW}}$  against their mean $(v_{\mathrm{RFID}} + v_{\mathrm{SW}})/2$ for each trial. The mean signed difference (bias), computed as $\overline{v_{\mathrm{RFID}} - v_{\mathrm{SW}}}$, is 0.061~$m/s$, indicating that the RFID system reports slightly higher speeds on average than the stopwatch. The 95\% limits of agreement (LOA) are  $[-0.070, 0.191]$~$m/s$, representing the range within which most inter-method differences are expected to fall.   
The positive bias is consistent with the known reaction-time delay in manual stopwatch operation, which tends to overestimate walking duration and therefore underestimate gait speed.  
Observing Fig.~\ref{fig:bland} reveals no proportional bias, as the differences do not increase systematically with higher mean gait speeds. This suggests that agreement between the RFID system and stopwatch measurement remains stable across the observed range of walking speeds.

\begin{figure}[t]
	\centering
	\includegraphics[width = 1 \linewidth]{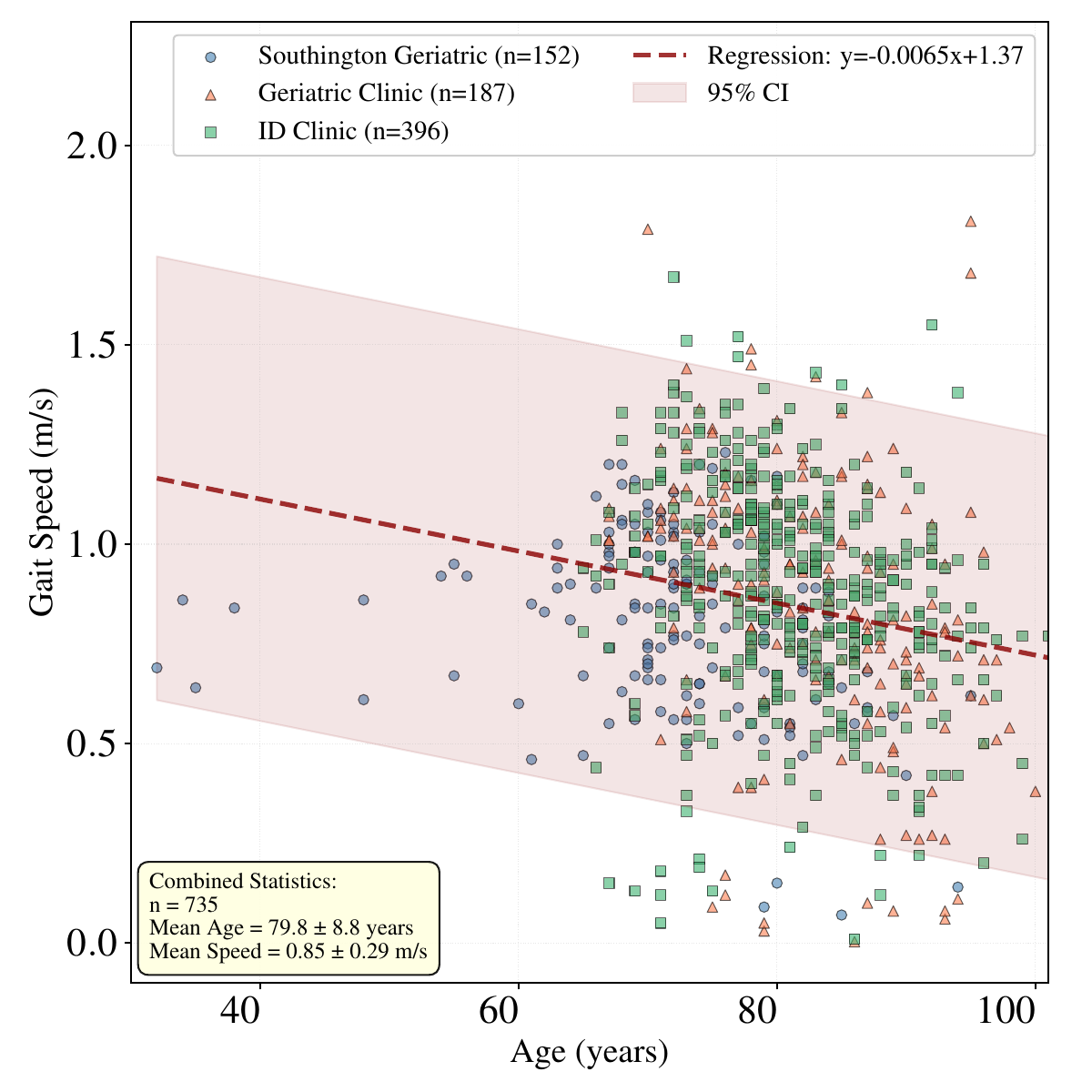}
	\caption{\small Gait Speed vs. Age Across Clinical Sites  }
	\label{fig:agespeed}
\end{figure}

Fig.~\ref{fig:agespeed} plots gait speed versus age for 734 measurements with recorded age information, with data points labeled by clinical site. Some valid gait-speed measurements lacked age documentation and are therefore excluded from this plot. We also removed seven abnormal gait-speed values outside the expected clinical range prior to fitting the regression. Despite differences in hallway geometry, local RF environments, and patient populations, all sites exhibit a consistent downward trend in gait speed with increasing age. A linear regression fit (dashed line) and its 95\% confidence interval (shaded region) capture this global relationship, demonstrating a clear association between increasing age and reduced gait speed in routine clinical measurements.

\section{Conclusion}
\label{sec:conclusion}
In this paper, we presented a fully passive, battery-free RFID-based gait-speed monitoring system designed for routine clinical deployment. The system combines a dual-antenna hardware configuration with a real-time, edge-based peak-detection algorithm that is robust to RSSI noise, plateau regions, and multiple local maxima, enabling automated gait-speed estimation without manual timing.
We evaluated the system through multi-site deployment across three UConn Health outpatient clinics, achieving an 87.7\% successful measurement rate across 966 trials and a mean absolute error of 0.064~$m/s$ compared with concurrent stopwatch timing.  Built entirely from commercial off-the-shelf components, the system offers a practical, low-cost alternative to instrumented walkways while preserving patient privacy and minimizing maintenance burden. The system has also been installed at two Atrium Health Wake Forest Baptist clinics, demonstrating portability across health systems.

In future work, we will focus on improving measurement success rates through continued data-driven refinement, expanding deployment to additional sites, and further validating clinical utility across broader patient populations and hallway configurations. We also plan to integrate gait-speed measurements directly into the Epic electronic health record system, enabling automated documentation and longitudinal tracking of patient mobility as part of routine clinical care. 

\newpage
\bibliography{bib}

@article{barry2018design,
  title={Design and validation of a radio-frequency identification-based device for routinely assessing gait speed in a geriatrics clinic},
  author={Barry, Lisa C and Hatchman, Laura and Fan, Zhaoyan and Guralnik, Jack M and Gao, Robert X and Kuchel, George A},
  journal={Journal of the American Geriatrics Society},
  volume={66},
  number={5},
  pages={982--986},
  year={2018},
  publisher={Wiley Online Library}
}

@article{barry2025taking,
  title={Taking the Next Step: Increasing Gait Speed Assessment in Primary Care},
  author={Barry, Lisa and Callahan, Kathryn and Wingood, Mariana and Fortinsky, Richard and Berg, Karina and Lin, Natong and Han, Song and Pajewski, Nicholas},
  journal={Innovation in Aging},
  volume={9},
  number={Supplement\_2},
  pages={igaf122--119},
  year={2025},
  publisher={Oxford University Press}
}

@article{middleton2015walking,
  title={Walking speed: the functional vital sign},
  author={Middleton, Addie and Fritz, Stacy L and Lusardi, Michelle},
  journal={Journal of aging and physical activity},
  volume={23},
  number={2},
  pages={314--322},
  year={2015},
  publisher={Human Kinetics, Inc.}
}

@article{studenski2011gait,
  title={Gait speed and survival in older adults},
  author={Studenski, Stephanie and Perera, Subashan and Patel, Kushang and Rosano, Caterina and Faulkner, Kimberly and Inzitari, Marco and Brach, Jennifer and Chandler, Julie and Cawthon, Peggy and Connor, Elizabeth Barrett and others},
  journal={JAMA},
  volume={305},
  number={1},
  pages={50--58},
  year={2011},
  publisher={American Medical Association}
}

@article{mejiacruz2021walking,
  title={Walking speed measurement technology: a review},
  author={MejiaCruz, Yohanna and Franco, Jean and Hainline, Garrett and Fritz, Stacy and Jiang, Zhaoshuo and Caicedo, Juan M and Davis, Benjamin and Hirth, Victor},
  journal={Current geriatrics reports},
  volume={10},
  number={1},
  pages={32--41},
  year={2021},
  publisher={Springer}
}

@article{briggs2020relationship,
  title={What is the relationship between orthostatic blood pressure and spatiotemporal gait in later life?},
  author={Briggs, Robert and Donoghue, Orna A and Carey, Daniel and O'Connell, Matthew DL and Newman, Louise and Kenny, Rose Anne},
  journal={Journal of the American Geriatrics Society},
  volume={68},
  number={6},
  pages={1286--1292},
  year={2020},
  publisher={Wiley Online Library}
}

@article{mcdonough2001validity,
  title={The validity and reliability of the GAITRite system's measurements: A preliminary evaluation},
  author={McDonough, Andrew L and Batavia, Mitchell and Chen, Fang C and Kwon, Soonjung and Ziai, James},
  journal={Archives of physical medicine and rehabilitation},
  volume={82},
  number={3},
  pages={419--425},
  year={2001},
  publisher={Elsevier}
}

@article{kirmizi2020effects,
  title={The effects of gait speed on plantar pressure variables in individuals with normal foot posture and flatfoot},
  author={Kirmizi, Muge and Sengul, Yesim S and Angin, Salih},
  journal={Acta of Bioengineering and Biomechanics},
  volume={22},
  number={3},
  pages={161--168},
  year={2020},
  publisher={Politechnika Wroc{\l}awska. Oficyna Wydawnicza Politechniki Wroc{\l}awskiej}
}

@inproceedings{boettcher2020dual,
  title={Dual-task gait assessment and machine learning for early-detection of cognitive decline},
  author={Boettcher, Lillian N and Hssayeni, Murtadha and Rosenfeld, Amie and Tolea, Magdalena I and Galvin, James E and Ghoraani, Behnaz},
  booktitle={2020 42nd Annual International Conference of the IEEE Engineering in Medicine \& Biology Society (EMBC)},
  pages={3204--3207},
  year={2020},
  organization={IEEE}
}

@article{lynall2017reliability,
  title={Reliability and validity of the protokinetics movement analysis software in measuring center of pressure during walking},
  author={Lynall, Robert C and Zukowski, Lisa A and Plummer, Prudence and Mihalik, Jason P},
  journal={Gait \& posture},
  volume={52},
  pages={308--311},
  year={2017},
  publisher={Elsevier}
}

@article{pham2024effects,
  title={The effects of cognition and vision while walking in younger and older adults},
  author={Pham, Trong and Suen, Meagan and Cho, Young-Hee and Krishnan, Vennila},
  journal={Sensors},
  volume={24},
  number={23},
  pages={7789},
  year={2024},
  publisher={MDPI}
}

@article{muro2014gait,
  title={Gait analysis methods: An overview of wearable and non-wearable systems, highlighting clinical applications},
  author={Muro-De-La-Herran, Alvaro and Garcia-Zapirain, Begonya and Mendez-Zorrilla, Amaia},
  journal={Sensors},
  volume={14},
  number={2},
  pages={3362--3394},
  year={2014},
  publisher={Molecular Diversity Preservation International (MDPI)}
}

@article{kitagawa2016estimation,
  title={Estimation of foot trajectory during human walking by a wearable inertial measurement unit mounted to the foot},
  author={Kitagawa, Naoki and Ogihara, Naomichi},
  journal={Gait \& posture},
  volume={45},
  pages={110--114},
  year={2016},
  publisher={Elsevier}
}

@article{washabaugh2017validity,
  title={Validity and repeatability of inertial measurement units for measuring gait parameters},
  author={Washabaugh, Edward P and Kalyanaraman, Tarun and Adamczyk, Peter G and Claflin, Edward S and Krishnan, Chandramouli},
  journal={Gait \& posture},
  volume={55},
  pages={87--93},
  year={2017},
  publisher={Elsevier}
}

@article{zhou2018hemodialysis,
  title={Hemodialysis impact on motor function beyond aging and diabetes—objectively assessing gait and balance by wearable technology},
  author={Zhou, He and Al-Ali, Fadwa and Rahemi, Hadi and Kulkarni, Nishat and Hamad, Abdullah and Ibrahim, Rania and Talal, Talal K and Najafi, Bijan},
  journal={Sensors},
  volume={18},
  number={11},
  pages={3939},
  year={2018},
  publisher={MDPI}
}

@article{keppler2019validity,
  title={Validity of accelerometry in step detection and gait speed measurement in orthogeriatric patients},
  author={Keppler, Alexander M and Nuritidinow, Timur and Mueller, Arne and Hoefling, Holger and Schieker, Matthias and Clay, Ieuan and B{\"o}cker, Wolfgang and F{\"u}rmetz, Julian},
  journal={PloS one},
  volume={14},
  number={8},
  pages={e0221732},
  year={2019},
  publisher={Public Library of Science San Francisco, CA USA}
}

@article{soltani2019real,
  title={Real-world gait speed estimation using wrist sensor: A personalized approach},
  author={Soltani, Abolfazl and Dejnabadi, Hooman and Savary, Martin and Aminian, Kamiar},
  journal={IEEE journal of biomedical and health informatics},
  volume={24},
  number={3},
  pages={658--668},
  year={2019},
  publisher={IEEE}
}

@article{farid2021feetme,
  title={FeetMe{\textregistered} Monitor-connected insoles are a valid and reliable alternative for the evaluation of gait speed after stroke},
  author={Farid, Leila and Jacobs, Damien and Do Santos, Johana and Simon, Olivier and Gracies, Jean-Michel and Hutin, Emilie},
  journal={Topics in stroke rehabilitation},
  volume={28},
  number={2},
  pages={127--134},
  year={2021},
  publisher={Taylor \& Francis}
}

@article{wu2020gaitway,
  title={GaitWay: Monitoring and recognizing gait speed through the walls},
  author={Wu, Chenshu and Zhang, Feng and Hu, Yuqian and Liu, KJ Ray},
  journal={IEEE Transactions on Mobile Computing},
  volume={20},
  number={6},
  pages={2186--2199},
  year={2020},
  publisher={IEEE}
}

@article{springer2016validity,
  title={Validity of the kinect for gait assessment: A focused review},
  author={Springer, Shmuel and Yogev Seligmann, Galit},
  journal={Sensors},
  volume={16},
  number={2},
  pages={194},
  year={2016},
  publisher={MDPI}
}

@article{geerse2015kinematic,
  title={Kinematic validation of a multi-Kinect v2 instrumented 10-meter walkway for quantitative gait assessments},
  author={Geerse, Daphne J and Coolen, Bert H and Roerdink, Melvyn},
  journal={PloS one},
  volume={10},
  number={10},
  pages={e0139913},
  year={2015},
  publisher={Public Library of Science San Francisco, CA USA}
}

@article{mazurek2024validation,
  title={A validation study demonstrating portable motion capture cameras accurately characterize gait metrics when compared to a pressure-sensitive walkway},
  author={Mazurek, Kevin A and Barnard, Leland and Botha, Hugo and Christianson, Teresa and Graff-Radford, Jonathan and Petersen, Ronald and Vemuri, Prashanthi and Windham, B Gwen and Jones, David T and Ali, Farwa},
  journal={Scientific reports},
  volume={14},
  number={1},
  pages={17464},
  year={2024},
  publisher={Nature Publishing Group UK London}
}

@inproceedings{li2022wivelo,
  title={Wivelo: Fine-grained walking velocity estimation for wi-fi passive tracking},
  author={Li, Chenning and Liu, Li and Cao, Zhichao and Zhang, Mi},
  booktitle={2022 19th Annual IEEE International Conference on Sensing, Communication, and Networking (SECON)},
  pages={172--180},
  year={2022},
  organization={IEEE}
}

@article{li2024wifi,
  title={WiFi-CSI difference paradigm: Achieving efficient doppler speed estimation for passive tracking},
  author={Li, Wenwei and Gao, Ruiyang and Xiong, Jie and Zhou, Jiarun and Wang, Leye and Mao, Xingjian and Yi, Enze and Zhang, Daqing},
  journal={Proceedings of the ACM on Interactive, Mobile, Wearable and Ubiquitous Technologies},
  volume={8},
  number={2},
  pages={1--29},
  year={2024},
  publisher={ACM New York, NY, USA}
}

@article{li2017indotrack,
  title={IndoTrack: Device-free indoor human tracking with commodity Wi-Fi},
  author={Li, Xiang and Zhang, Daqing and Lv, Qin and Xiong, Jie and Li, Shengjie and Zhang, Yue and Mei, Hong},
  journal={Proceedings of the ACM on Interactive, Mobile, Wearable and Ubiquitous Technologies},
  volume={1},
  number={3},
  pages={1--22},
  year={2017},
  publisher={ACM New York, NY, USA}
}

@inproceedings{qian2017widar,
  title={Widar: Decimeter-level passive tracking via velocity monitoring with commodity Wi-Fi},
  author={Qian, Kun and Wu, Chenshu and Yang, Zheng and Liu, Yunhao and Jamieson, Kyle},
  booktitle={Proceedings of the 18th ACM international symposium on mobile ad hoc networking and computing},
  pages={1--10},
  year={2017}
}

@inproceedings{qian2018widar2,
  title={Widar2. 0: Passive human tracking with a single Wi-Fi link},
  author={Qian, Kun and Wu, Chenshu and Zhang, Yi and Zhang, Guidong and Yang, Zheng and Liu, Yunhao},
  booktitle={Proceedings of the 16th annual international conference on mobile systems, applications, and services},
  pages={350--361},
  year={2018}
}

@article{pua2025gait,
  title={Gait speed assessment in confined spaces: Development of a novel automated 4-m static-start test to measure dynamic-start gait speed},
  author={Pua, Yong-Hao and Clark, Ross Allan and Tay, Laura and Ng, Yee-Sien and Poh, Jaylyn Tze-Theng and Ibrahim, Salma Bte Md and Cheong, Wai-Chye and Tan, Hong-Han and Thumboo, Julian},
  journal={Geriatrics \& Gerontology International},
  volume={25},
  number={3},
  pages={449--453},
  year={2025},
  publisher={Wiley Online Library}
}

@article{sansano2022continuous,
  title={Continuous non-invasive assessment of Gait speed through Bluetooth low energy},
  author={Sansano-Sansano, Emilio and Montoliu, Ra{\'u}l and Belmonte-Fern{\'a}ndez, {\'O}scar and Aranda, Fernando J and {\'A}lvarez, Fernando J},
  journal={IEEE Sensors Journal},
  volume={22},
  number={8},
  pages={8183--8195},
  year={2022},
  publisher={IEEE}
}

@article{gurbuz2024overview,
  title={Overview of radar-based gait parameter estimation techniques for fall risk assessment},
  author={Gurbuz, Sevgi Z and Rahman, Mohammad Mahbubur and Bassiri, Zahra and Martelli, Dario},
  journal={IEEE Open Journal of Engineering in Medicine and Biology},
  volume={5},
  pages={735--749},
  year={2024},
  publisher={IEEE}
}

@article{wang2024extraction,
  title={Extraction and validation of biomechanical gait parameters with contactless FMCW radar},
  author={Wang, Linyu and Ni, Zhongfei and Huang, Binke},
  journal={Sensors},
  volume={24},
  number={13},
  pages={4184},
  year={2024},
  publisher={MDPI}
}

@inproceedings{huang2022rfid,
  title={RFID based Gait Speed Measurement using Doppler Shift},
  author={Huang, Kai and Zhang, Jintao and Chu, Yicheng and Ma, Yongtao},
  booktitle={2022 IEEE 12th International Conference on RFID Technology and Applications (RFID-TA)},
  pages={13--16},
  year={2022},
  organization={IEEE}
}

@article{mo2018study,
  title={A Study of Walking Speed Measurement for Elderly Health Assessment Using Ultrahigh-frequency Radio-frequency Identification Tags.},
  author={Mo, Lingfei and Li, Chenyang and Huang, Hualin and Dong, Yaxuan},
  journal={Sensors \& Materials},
  volume={30},
  year={2018}
}

@article{gay2021novel,
  title={Novel use of radio frequency identification (RFID) provides a valid measure of indoor stair-based physical activity},
  author={Gay, Jennifer L and Carmichael, Kaitlyn E and LaFlamme, Chantal C and O'Connor, Patrick J},
  journal={Applied Ergonomics},
  volume={95},
  pages={103431},
  year={2021},
  publisher={Elsevier}
}

@techreport{epc_standard,
  author      = {{GS1}},
  title       = {{EPC\textregistered\ Radio-Frequency Identity Generation-2 UHF RFID Standard}},
  institution = {GS1},
  type        = {Standard},
  number      = {Version 3},
  year        = {2024},
  month       = jan,
  url         = {https://www.gs1.org/standards/rfid/uhf-air-interface-protocol}
}

@misc{impinj_reader_modes_2022,
  author       = {{James Skinner}},
  title        = {{Reader Modes (RF Modes) Made Easy}},
  year         = {2022},
  month        = apr,
  day          = {05},
  organization = {Impinj, Inc.},
  url          = {https://support.impinj.com/hc/en-us/articles/360000046899-Reader-Modes-RF-Modes-Made-Easy}
}

@misc{vulcan_s9028pc_antenna,
  author       = {{Atlas RFID Store / Vulcan RFID}},
  title        = {{Vulcan RFID S9028PC RFID Panel Antenna Datasheet}},
  year         = {2025},
  note         = {\url{https://rfid.atlasrfidstore.com/hubfs/1_Tech_Spec_Sheets/Vulcan%20RFID/ATLAS%20Vulcan%20RFID%20S9028PC%20RFID%20Panel%20Antenna.pdf}}

}

@article{maggio2016instrumental,
  title={Instrumental and non-instrumental evaluation of 4-meter walking speed in older individuals},
  author={Maggio, Marcello and Ceda, Gian Paolo and Ticinesi, Andrea and De Vita, Francesca and Gelmini, Giovanni and Costantino, Cosimo and Meschi, Tiziana and Kressig, Reto W and Cesari, Matteo and Fabi, Massimo and others},
  journal={PloS one},
  volume={11},
  number={4},
  pages={e0153583},
  year={2016},
  publisher={Public Library of Science San Francisco, CA USA}
}

@article{bland2010statistical,
  title={Statistical methods for assessing agreement between two methods of clinical measurement},
  author={Bland, J Martin and Altman, Douglas G},
  journal={International journal of nursing studies},
  volume={47},
  number={8},
  pages={931--936},
  year={2010},
  publisher={Elsevier}
}
\end{document}